\documentclass[journal,transmag]{IEEEtran}
\usepackage[top=0.64in, bottom=0.64in, left=0.64in, right=0.64in]{geometry}
\IEEEoverridecommandlockouts
\usepackage{booktabs}
\usepackage{setspace}
\usepackage{multirow}
\usepackage{color}
\usepackage{float}
\usepackage{cite}
\usepackage{enumitem}  
\usepackage[utf8]{inputenc}
\usepackage{lmodern}
\usepackage{tabularx}
\usepackage{csquotes}
\usepackage[justification=centering]{caption}
\usepackage{blindtext}
\usepackage{wrapfig}
\usepackage{graphicx}
\usepackage[bottom]{footmisc}
\DeclareGraphicsExtensions{.eps,.pdf,.gif,.png,.jpg}
\usepackage{amssymb}
\usepackage{amsthm}
\usepackage{array}
\usepackage{algorithm,algorithmic}
\usepackage{booktabs}
\usepackage{graphicx}
\usepackage{epstopdf}\usepackage[utf8]{inputenc}
\usepackage[hyphens,spaces,obeyspaces]{url}
\usepackage[english]{babel}
\usepackage{verbatim} 
\usepackage{lineno,nohyperref}
\usepackage{epstopdf} 
\modulolinenumbers[5]

\RequirePackage{filecontents}
\usepackage{ragged2e}
\usepackage{url} 
\usepackage[T1]{fontenc}
\usepackage[utf8]{inputenc}
\usepackage{microtype} 

\newcommand{\vect}[1]{\boldsymbol{#1}}
\newcommand*{\email}[1]{#1}
\PassOptionsToPackage{hyphens}{url}
\newtheoremstyle{mystyle}
{}
{}
{\itshape}
{}
{\bfseries}
{.}
{ }
{\thmname{#1}\thmnumber{ #2}\thmnote{ (#3)}}

\theoremstyle{mystyle}

\newcounter{subassumption}[asu]

\makeatletter
\renewcommand{\p@subassumption}{\theasu}
\makeatother
\usepackage{amsthm}
\usepackage{xpatch,amsthm}
\makeatletter
\xpatchcmd{\@thm}{\fontseries\mddefault\upshape}{}{}{} 
\makeatother

\makeatother
\usepackage{cite}
\usepackage{amsmath,amssymb,amsfonts}
\usepackage{algorithmic}
\usepackage{graphicx}
\usepackage{textcomp}
\usepackage{xcolor}
\def\BibTeX{{\rm B\kern-.05em{\sc i\kern-.025em b}\kern-.08em
    T\kern-.1667em\lower.7ex\hbox{E}\kern-.125emX}}

\begin{document}

\title{ Federated Learning Assisted Deep Q-Learning for  Joint Task Offloading  and Fronthaul Segment Routing in Open RAN 
\thanks{}
}
 \author{Anselme~Ndikumana,~
	Kim~Khoa~Nguyen,~\IEEEmembership{Senior~Member,~IEEE,~}\\
	and~Mohamed~Cheriet,~\IEEEmembership{Senior~Member,~IEEE,~}
	\IEEEcompsocitemizethanks{
		\IEEEcompsocthanksitem Anselme Ndikumana, Kim Khoa Nguyen, and Mohamed Cheriet (Corresponding author) are  with Synchromedia Lab, École de
		Technologie Supérieure, Université du Québec, 1100 Notre-Dame St W, Montreal, QC H3C 1K3, Canada, (E-mail: \email{anselme.ndikumana.1@ens.etsmtl.ca; kim-khoa.nguyen@etsmtl.ca; Mohamed.Cheriet@etsmtl.ca}).
	}}
\maketitle

\begin{abstract}
Offloading computation-intensive tasks to edge clouds has become an efficient way to support resource constraint edge devices. However, task offloading delay is an issue largely due to the networks with limited capacities between edge clouds and edge devices. In this paper, we consider task offloading in Open Radio Access Network (O-RAN), which is a new 5G RAN architecture allowing Open Central Unit (O-CU) to be co-located with Open Distributed Unit (DU) at the edge cloud for low-latency services. O-RAN  relies on fronthaul network to connect  O-RAN Radio Units (O-RUs) and edge clouds that host O-DUs. Consequently, tasks are offloaded onto the edge clouds via wireless and fronthaul networks \cite{10045045}, which requires routing. Since edge clouds do not have the same available computation resources and tasks' computation deadlines are different, we need a task distribution approach to multiple edge clouds. Prior work has never addressed this joint problem of task offloading, fronthaul routing, and edge computing. To this end, using segment routing, O-RAN intelligent controllers, and multiple edge clouds, we formulate an optimization problem to minimize offloading, fronthaul routing, and computation delays in O-RAN. To determine the solution of this NP-hard problem, we use Deep Q-Learning assisted by federated learning with a reward function that reduces the Cost of Delay (CoD). The simulation results show that our solution maximizes the reward in minimizing CoD.
\end{abstract}

\section{Introduction}
\label{sec:introduction}
 By the year 2025, there will be 34.2 billion edge devices, including 21.5 billion IoT devices \cite{IoTAnalytics}. Consequently,  edge devices will be anywhere, anytime, and connected to anything. Therefore, it will be not only people who generate data but also machines/things \cite{ndikumana2019joint}. However, edge devices have limited resources such as memory, CPU, and energy. Offloading computation-intensive tasks to edge clouds helps resource-constrained edge devices address this issue. However,  networks between edge clouds and edge devices critically impact offloading delay. 
 \begin{figure}[t]
 \centering
 \includegraphics[width=1.0\columnwidth]{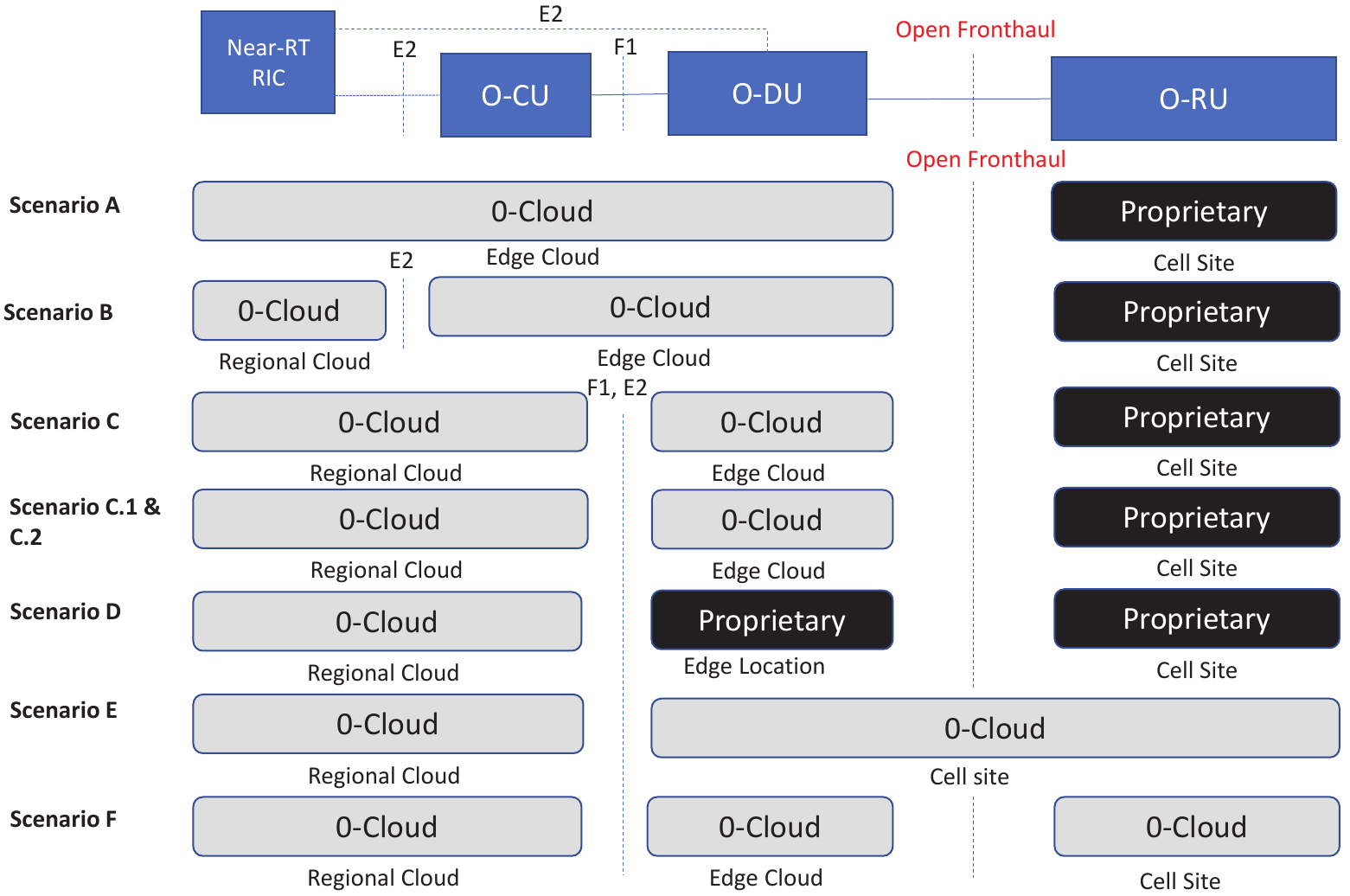}
 \caption{O-RAN deployment scenarios \cite{allianceORANUseCases}.}
 \label{fig:o-ran}
 \end{figure}
 
To provide edge devices with low latency and high data rates, the 5G Radio Access Network (RAN) has experienced some transformations to increase deployment flexibility and network dynamics. The recent RAN transformation is Open Radio Access Network (O-RAN) architecture \cite{ArchitectureDescription}. In O-RAN deployment scenarios shown in Fig. \ref{fig:o-ran} and described in \cite{allianceORANUseCases}, O-DU and O-CU, where O stands for Open, can be modular base station software stacks on off-the-shelf server hardware that different vendors can supply.  A low latency offloading service may require O-RAN deployment scenario A, where O-CU can be co-located with the O-DU at the edge cloud (e.g., a telecom room close to the edge devices). In scenario A, the edge cloud has O-Cloud that hosts O-RAN Central Unit Control Plane (O-CU-CP) and O-RAN Central Unit User Plane (O-CU-UP), O-DU, and Near-Real Time RAN Intelligent Controller (Near-RT RIC). Nearly real-time RAN resources and elements can be optimized using Machine Learning (ML) algorithms implemented in Near-RT RIC. Also, O-RAN has a Non-Real Time RAN Intelligent Controller (Non-RT RIC) that enables Machine Learning (ML) functionalities for policy-based guidance of applications and features. Using O-RAN, the tasks are offloaded onto the edge cloud via a wireless network between edge devices and O-RUs, and the fronthaul network linking O-RUs with the edge cloud that hosts O-DU \cite{ndikumana2022age}. Ideally, offloading should be based on a reliable fronthaul and wireless connection between edge devices and edge clouds. Unfortunately, it is not always the case in reality. Offloading traffic pressures and strict delay requirements for time-sensitive applications can significantly challenge wireless and fronthaul networks. 
 
Currently, there are many fronthaul transport technologies described in \cite{alimi2017toward} to reach edge clouds. Some of these technologies are microwave,  Passive Optical Network (PON),  Wavelength-Division Multiplexing (WDM) PON, Coarse-WDM (CWDM) PON, Dense-WDM (DWDM), and Ethernet. Ethernet-based fronthaul network is a  lower-cost solution that can reduce Capital Expenditure (CapEx) and Operating Expenses (OpEx) compared to other technologies \cite{alimi2017toward}. Therefore, this paper considers Time-Sensitive Networking (TSN) \cite{farkas2018time} for Fronthaul as an Ethernet-based solution. IEEE 802.1CM \cite{2020standard} standard and Common Public Radio Interface (eCPRI) \cite{perez20195g} allow to connect O-RUs to O-DUs using  a packet network. Fronthaul traffic over the packet network enables switched connectivity between O-RUs and O-DUs. Therefore,  we can route fronthaul traffic to edge clouds using multiple paths and hops. A class of low-latency 5G applications requires a fronthaul delay of  $100$ µs for the user plane traffic \cite{Viavi}. To meet this requirement, we need a routing approach that simplifies the existing IP/Multi-Protocol Label Switching (MPLS) based strategies to reduce fronthaul latency. We consider Segment Routing (SR) \cite{tong2021sdn} in the fronthaul network as a routing solution because it simplifies the control plane by removing the need for a per-flow state to be maintained at each node in MPLS. In other words, in SR, the per-flow state is only maintained at the ingress node of an SR domain. Still,  there are  many critical challenges for data offloading, fronthaul routing, and edge computing that have never been addressed in the literature, such as:
\begin{itemize}
	\item 
	The problem of task offloading, fronthaul routing, and edge computing should be addressed jointly to satisfy task computation deadlines.
	\item 
	In forwarding decisions, TSN bridges for the fronthaul network use Ethernet header contents, not IP addresses. Therefore, we need a network approach that extends layer $2$ as a network overlay for TSN fronthaul routing.
	\item 
	When each edge cloud operates independently, required resources for offloaded computation-intensive tasks to the edge cloud may exceed available computation resources. Therefore, task distribution to multiple edge clouds should be considered.
	\item 
	Computation tasks have different deadlines. Also, edge clouds do not have the same available computation resources. Therefore, we need a task distribution approach to multiple edge clouds to meet computation deadlines. 
\end{itemize}

In this work, we opt for O-RAN and take advantage of O-RAN intelligent controllers to tackle the abovementioned challenges of task offloading, fronthaul routing, and edge computing.
However, O-RAN is not restrictive; other RAT technologies considering the 7-2x split option  (fronthaul between DU and RU can be applied).   
The main contributions of this paper are summarized as follows:
\begin{itemize}
	\item 
	We propose offloading approach for edge devices. The proposed approach enables edge devices with insufficient computation resources to balance the costs between keeping the computational task until the resources become available for local computation and offloading tasks to the edge cloud. 
	\item
	We propose O-RAN fronthaul routing approach using Near-RT RIC to route offloaded tasks to edge clouds. Since fronthaul TSN bridges use the Ethernet header contents, not the IP addresses, we consider Virtual Extensible LAN protocol (VXLAN) \cite{carthern2021introduction}  to extend layer two connectivity as a network overlay. Then, we apply SR in the fronthaul network to route offloaded tasks to multiple edge clouds. To the best of our knowledge, this research is the first that leverages O-RAN controllers, SR, and VXLAN in a joint task offloading, fronthaul routing, and edge computing problem.
	\item 
	We propose an edge cloud computing approach that enables edge cloud with insufficient resources to request computation support to its neighbor edge clouds or regional cloud through redirecting  offloaded tasks. 
	\item 
	We formulate an optimization problem to minimize offloading, fronthaul routing, and computation delay. We convert the proposed NP-hard problem to the reward function for a dynamic offloading environment. Then, we design a Deep Q-Learning approach, assisted by federated learning, to maximize the reward function by reducing CoD.
\end{itemize}

As related work, task offloading in wireless networks \cite{mustafa2021joint, ndikumana2017collaborative,ndikumana2019joint, ndikumana2019intelligentedge, ndikumana2019self} has gained significant attention in research communities compared to offloading in fronthaul networks. 
Considering wireless and fronthaul networks between edge devices and edge clouds, the authors in \cite{wu2016computing} proposed an offloading approach in Cloud Radio Access Network (C-RAN), where the mobile device can change the offloading strategy to reduce fronthaul traffic. In \cite{kaneva2021multi} authors propose multi-hop fronthaul offloading in C-RAN and compare multi-hops with single-hop communication. For fronthaul routing, the authors in \cite{nakayama2018low}  proposed a lower latency scheme in C-RAN that enables the selection of a set of paths that minimizes delays from the preselected k shortest paths. The proposed C-RAN-based approaches consider fronthaul between Remote
Radio Unit (RRU) and Baseband Processing Unit (BBU). However, we have fronthaul between O-RUs and O-DUs and middlehaul between DUs and CUs in O-RAN. Since our proposal considers O-DU and O-CU to be hosted at the same edge cloud in O-RAN, the middlehaul network is outside this paper's scope.
To apply machine learning in data offloading,  the authors in \cite{park2020communication} highlight that exchanging raw data may slow down wireless communication services. Therefore, the authors in \cite{li2020federated} emphasized that sharing only machine learning parameters of federated learning without sharing raw data can be an appropriate solution to address this issue. The authors in \cite{xiong2020intelligent} proposed intelligent task offloading approach that uses a federated Q-learning method to minimize the probabilities of data offloading failure by considering communication and computing budgets. However, the authors did not consider fronthaul network in offloading data to the edge cloud. In  \cite{chen2020intelligent}, the authors applied a deep Monte Calor tree search in data offloading. Their proposed approach enables an agent to observe the network environment and decide the offloading actions. To address the issue of limited resource sharing in data offloading, the authors in \cite{chen2020computation} formulated a computation offloading problem using a multi-agent Markov decision process in multi-access edge computing. However, the previous related works have never addressed the issue of joint task offloading,  fronthaul routing,  and edge computing.

The rest of this paper is structured as follows: we present our system model in Section \ref{sec:system-model}, while Section \ref{sec:DeepLearningOptimizationl} describes in detail our task offloading, fronthaul routing, and cloud computation. We present our problem formulation and solution in Section \ref{sec:Problem_Formulation_Solution}. Section \ref{sec: SimulationResultsAnalysis} presents our performance evaluation.  We conclude the paper in Section \ref{sec:Conclusion}.

\section{System model}
\label{sec:system-model}

\begin{figure}[t]
	\centering
	\includegraphics[width=1.0\columnwidth]{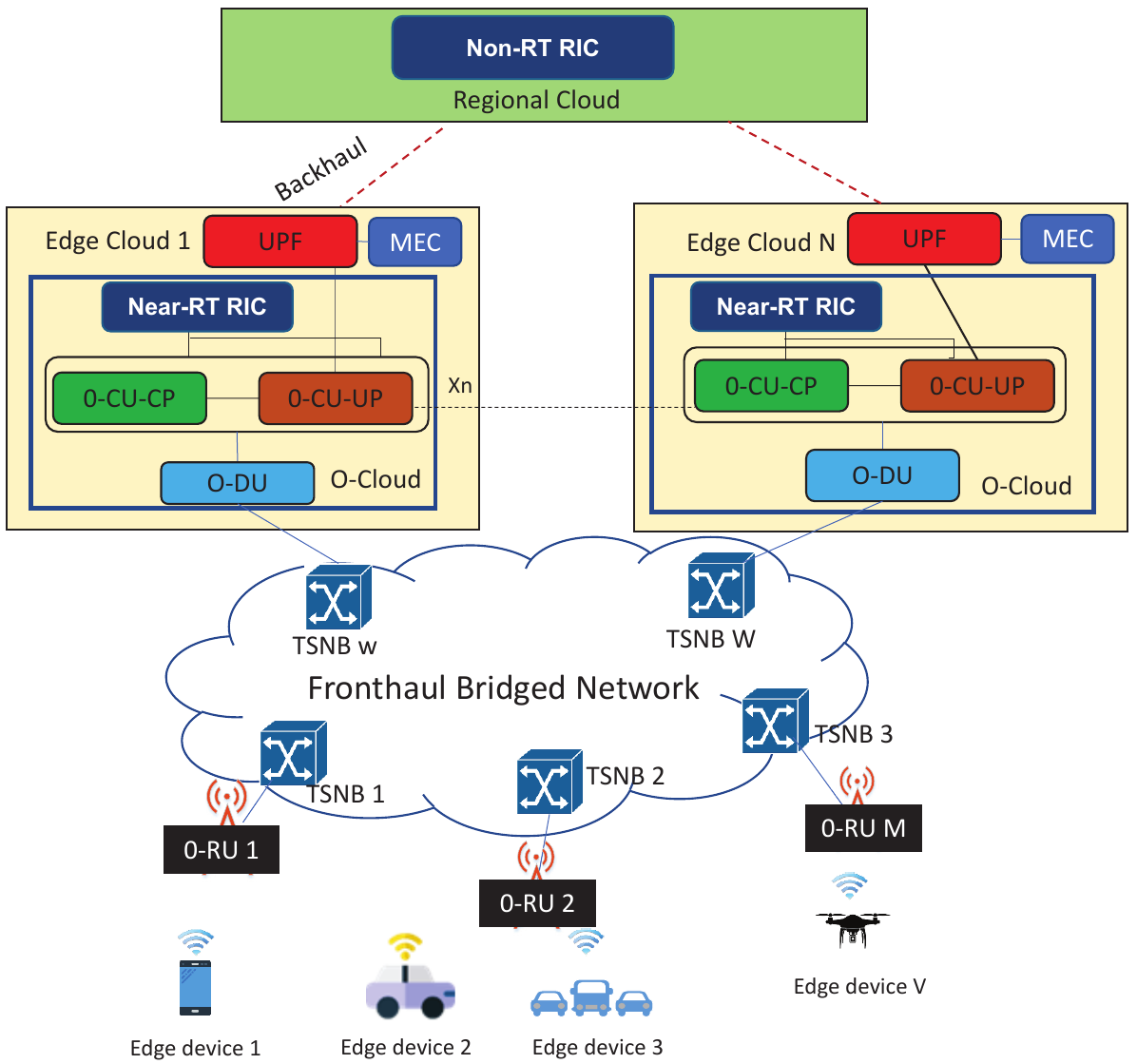}
	\caption{Illustration of our system model.}
	\label{fig:SystemModel}
\end{figure}
\begin{table}[t]
		\caption{Summary of key notations.}
		\label{tab:table1}
		\begin{tabular}{ll}
				\toprule
				Notation & Definition\\
				\midrule
				$\mathcal{N}$ & Set of Edge Clouds (ECs), $|\mathcal{N}|= N$\\
				$\mathcal{M}$ & Set of O-RUs,  $|\mathcal{M}|= M$\\
				$\mathcal{V}$ & Set of edge devices,  $|\mathcal{V}|= V$\\
				$\mathcal{W}$ & Set of  fronthaul TSNBs, $|\mathcal{W}|= W$\\
				$\Gamma_{v}$ & Computation task of edge device $v \in \mathcal{V}$\\
				$\tau^\textrm{loc}_{v}$ & Local computational delay of device $v$\\
				$\tau^\textrm{off}_{v}$ & Offloading delay of device $v$\\ 
				$\mu_{v}$& Status parameter of device $v$\\ 
			    $\mathcal{E}$ & Set of fronthaul links, $|\mathcal{E}|= E$\\
				$\chi_n$ & Computation capability of EC	$n \in \mathcal{N}$\\
				$\chi_v$ & Computation resource of  device $v \in \mathcal{V}$\\
				$x_v^{m \rightarrow n}$ & Offloading variable of  device $v \in \mathcal{V}$\\
				$\omega^m_v$ & Wireless capacity between device $v$\\
				& and EC $m$\\
				$\omega^j_i$	& Fronthaul capacity of link $e_{i,j}$\\ 
				$\omega^q_n$ 	& Link capacity between EC $n$ and  EC $q$\\
				$\omega^{RC}_n$ & Backhaul capacity between EC $n$ and RC\\
				$\mathcal{R}$ &  Reward\\
				$ \mathcal{A}$ & Action space \\
				$\mathcal{S}$& State space \\ 
				$\mathcal{P}$&Transition probability matrix\\ 
				\bottomrule
			\end{tabular}
	\end{table}

The system model of our joint task offloading, fronthaul routing, and cloud computation approach is depicted in Fig. \ref{fig:SystemModel}. For easy visualization 
of the system model, we omit some interfaces in Fig. \ref{fig:SystemModel}.

In the system model, we consider $\mathcal{V}$ as a set of edge devices.  Each edge device $ v \in \mathcal{V}$ has computation-intensive tasks that need to use computation resources, such as on-device machine learning and Extended Reality (it combines Virtual Reality and Augmented Reality \cite{rantakokko2022data}). For example,  in  XR, edge devices participating in crowd-sensing \cite{nguyen2021mobile} can sense their environment, compute sensed data, and create virtual environment. However, if the edge device does not have resources, it can send tasks and data to the edge cloud for computation and creating the virtual environment.  We define a task $\Gamma_{v} = (d_v,\tilde{\tau}_{v}, \tilde{z}_{v})$, where $d_v$ is the size of computation input data from edge device $v$ in terms of bits, $\tilde{\tau}_{v}$ is the task computation deadline, and $\tilde{z}_{v}$ is the computation workload or intensity in terms of CPU cycles per bit. Each edge device $ v \in \mathcal{V}$ has computation capability $\chi_v$. Offloading happens when the edge device does not have enough computation resources, and it can not hold computational tasks until the resource becomes available.  

Task offloading requires communication resources. Therefore, each edge device $ v \in \mathcal{V}$ is connected to O-RU $m  \in \mathcal{M}$  via a wireless link of capacity $\omega^m_v$.
Here, we denote $\mathcal{M}$ as a set of O-RUs. Once offloaded tasks reach O-RUs, the O-RUs forward the task to O-DUs using Fronthaul Bridged Network (FBN) of multiple paths. We model FBN
as  graph $\mathcal{G}=(\mathcal{W},\mathcal{E})$, where $\mathcal{W}$ is the set of Time-Sensitive Networking Bridges (TSNBs)  and $\mathcal{E}$ is the set of links.  In FBN, each O-RU $m$ is connected to ingress TSNB, and each O-DU is connected to egress TSNB. We denote $\mathcal{N}$ as a set of Edge Clouds (ECs) that host O-DUs.  We use the terms \enquote{EC} and \enquote{O-DU} interchangeably. O-DU $ n \in \mathcal{N}$ means the O-DU hosted at EC $n$.   To route offloaded tasks using multiple paths, we use SR in FBN, where Near-RT RIC at EC controls fronthaul SR. To implement SR described in Section \ref{sec:SegmentroutingFronthaulBridgedNetwork}, we assume that VXLAN  is applied in FBN. We choose VXLAN \cite{carthern2021introduction} over Virtual LAN (VLAN) because   VXLAN uses a VXLAN network identifier of  $24$ bits, while  VLAN has a network identifier of $12$ bits. Therefore,  VLAN can be scaled up to $4000$ VLANs, while VXLAN  can be scaled up to $16$ million VXLANs segments. Combining SR and VXLAN can help the network to handle massive fronthaul traffic offloading that needs to be routed to multiple ECs.  In implementing SR using VXLAN, we assume Near-RT RIC knows the FBN topology and can communicate with all TSNBs. The Near-RT RIC records traffic matrix. With the MEC server's help, traffic matrix, and network topology information, the Near-RT RIC  has the segment paths for each source-destination pair in the FBN.

Each EC has an MEC server, User Plane Function (UPF), and O-Cloud to improve reliability and lower latency in data offloading.    Once the tasks reach O-DU via egress TSNB, the O-DU sends them to the MEC server accessible via O-CU-UP and UPF for computation. Each MEC $n \in \mathcal{N}$ has a computational resource of capacity $\chi_n$ that can be allocated to edge devices. Here, unless stated otherwise, we use the terms \enquote{EC} and \enquote{MEC} interchangeably. MEC $ n \in \mathcal{N}$ means the MEC hosted at EC $n$. Here, we assume ECs can exchange application-level data using  Xn interface \cite{huang2020prospect}. Furthermore, each EC $n \in \mathcal{N}$ has access to the Regional Cloud (RC) via a wired backhaul of capacity $\omega^{RC}_n$. When computation resources are unavailable in ECs, the tasks will be offloaded to the RC in the worst-case scenario. We denote  $\chi_{RC}$ as the computation
capacity of the RC. Each RC hosts Non-RT RIC. Unless stated otherwise, we use the terms \enquote{RC}, and \enquote{Non-RT RIC} interchangeably.

\section{Task Offloading, Fronthaul Routing, and Cloud Computation}
\label{sec:DeepLearningOptimizationl}

This section discusses our task offloading approach that enables edge devices to balance the costs of local computation and offloading tasks to the edge cloud. Furthermore, we present the models for wireless communication and fronthaul routing that enable offloaded tasks to reach edge clouds. We conclude the section with a computation model for edge and regional clouds.

\subsection{ Computation at Edge Devices}
\label{subsec:massiveData_handlingDevice} 

Each edge device $ v \in \mathcal{V}$ has an application that generates computation task $\Gamma_{v}$. Computing task $\Gamma_{v}$ at edge device $v$ requires CPU computation resources. By using computation resource $\chi_v$, the execution latency for task $\Gamma_{v}$ at edge device $v$ is given by:
	\begin{equation}
		\label{eq:compution_time}
		\tau_v=\frac{d_v\tilde{z}_{v} }{\chi_v}.
\end{equation}

When $\tau_v> \tilde{\tau}_{v}$, or $ \tilde{z}_{v}>\chi_v$, edge  device $v$ does not have enough resources or cannot meet the computation deadline. Therefore, the edge device can hold on the computational task until the resources become available for local computation or offload the task to the edge cloud. To handle such a situation, we define the edge device status parameter $\mu_{v}\in \{0,1\}$, where $\mu_{v}$ is expressed as follows:
	\begin{equation}
		\setlength{\jot}{10pt}
		\mu_{v} =
		\begin{cases}
			0,\; \text{if $\tau_v> \tilde{\tau}_{v}$, or $ \tilde{z}_{v}>\chi_v$,}\\
			1, \;\text{otherwise.}
		\end{cases}
\end{equation}
Based on $\mu_{v}\in \{0,1\}$, the total local execution time $\tau^\textrm{loc}_{v}$ of task $\Gamma_{v}$ at edge device $v$ becomes:
\begin{equation}
	\setlength{\jot}{10pt}
	\label{eq:local}
	\tau^\textrm{loc}_{v} =
	\begin{cases}
		\tau_v\;, \text{if $\mu_{v}=1$, and  $x_v^{m \rightarrow n}=0$},\\
		\tau_v+ \varphi_v\;, \text{if $\mu_{v}=0$, and  $x_v^{m \rightarrow n}=0$},\\
		0, \; \text{if $\mu_{v}=0$ and  $x_v^{m \rightarrow n}=1$},\\
	\end{cases}
\end{equation}
where $\varphi_v$ is the average hold time for task $\Gamma_{v}$ until it is locally computed at edge device $v$. The $\varphi_v$ can be considered as a time for charging the battery or finishing ongoing computation of other tasks than $\Gamma_{v}$. When edge device $v$  cannot hold on computational task, it can offload task to the edge cloud. Therefore, we define $ x_v^{m \rightarrow n}\in \{0,1\}$ as an offloading decision variable, where $ x_v^{m \rightarrow n}$ is given by:
\begin{equation}
	\setlength{\jot}{10pt}
	x_v^{m \rightarrow n}=
	\begin{cases}
		1,\; \text{if $\Gamma_{v}$ is offloaded from edge device $v$  to }\\ \text{\; \; EC $n$ via O-RU $m$}\\
		0, \;\text{otherwise.}
	\end{cases}
\end{equation}
\subsection{Task Offloading in Wireless Networks}
\label{sec:TaskOffloadingFronthaulRouting}
Offloading a task from the edge device $v \in \mathcal{V}$ to  EC $n \in \mathcal{N}$  via O-RU $m  \in \mathcal{M}$ requires wireless communication  between  each edge device $v$  and O-RU $m$. The spectrum efficiency (as described in \cite{zhou2017resource}) for edge device $v$ is given by:
\begin{equation}
	\label{eq:SINR}
	\begin{aligned}
		\gamma^m_v = \log_2\left(1 + \frac{\rho_v |G^m_v|^2}{\sigma_v^2}\right),  \;\forall v \in \mathcal{V},\; m  \in \mathcal{M}.
	\end{aligned}
\end{equation}
Here, $\rho_v$ is the transmission power of edge device, $|G^m_v|^2$ is the channel gain between edge device $v$ and O-RU $m$, and $\sigma_v^2$ is the power of the Gaussian noise at edge device $v$. The instantaneous data rate for edge device $v$ is expressed as:
\begin{equation}
	B^m_v= x_v^{m \rightarrow n}b_v^m \omega^m_v\gamma^m_v, \forall v \in \mathcal{V},\; m\in \mathcal{M}, 
	\label{eq:instantaneous_data}
\end{equation} 
where each edge device $v$  is allocated a fraction $b_v^m$ ($ 0\leq b_v^m \leq 1$) of bandwidth $\omega^m_v$. We assume that the spectrum of the mobile network operator is orthogonal, and there is no interference among the edge devices. Furthermore, we assume that the demand of edge devices for task offloading will only be accepted if there are enough spectrum resources to satisfy its demand. Based on the instantaneous data rate, the transmission delay for offloading a task from edge device $v$ to  EC $n$ is expressed as:
\begin{equation}
	\tau^{v \rightarrow m}=x_v^{m \rightarrow n}\frac{ d_v}{B^m_v}, \; \forall v \in \mathcal{V} , \;m  \in \mathcal{M}.
\end{equation}

\subsection{Task Offloading in Fronthaul Bridged Network}
\label{sec:SegmentroutingFronthaulBridgedNetwork}
When the offloaded tasks reach O-RUs, the O-RUs forward them to O-DUs using FBN. Since O-DUs are hosted at ECs, the offloaded tasks can reach ECs using multiple fronthaul paths. To route the offloaded tasks to O-DU, we use SR described below.
\subsubsection{Overview of Segment Routing}
The existing fronthaul routing in \cite{nakayama2018low} is based on the shortest path algorithm in C-RAN. Here, we use SR \cite{tong2021sdn} in O-RAN as a source routing approach because it overcomes the MPLS Traffic-Engineering (MPLS-TE) per-flow state that needs to be maintained in each network node to support traffic-engineered paths in IP backbones. SR improves  MPLS-TE in labeling,  where SR does not require configuring forwarding tables in each node along the transmission path. SR includes the route in the packet header at the ingress node. In other words, SR adds a list of hops in the packet header as a route. 

\begin{figure}[t]
	\centering
	\includegraphics[width=1.0\columnwidth]{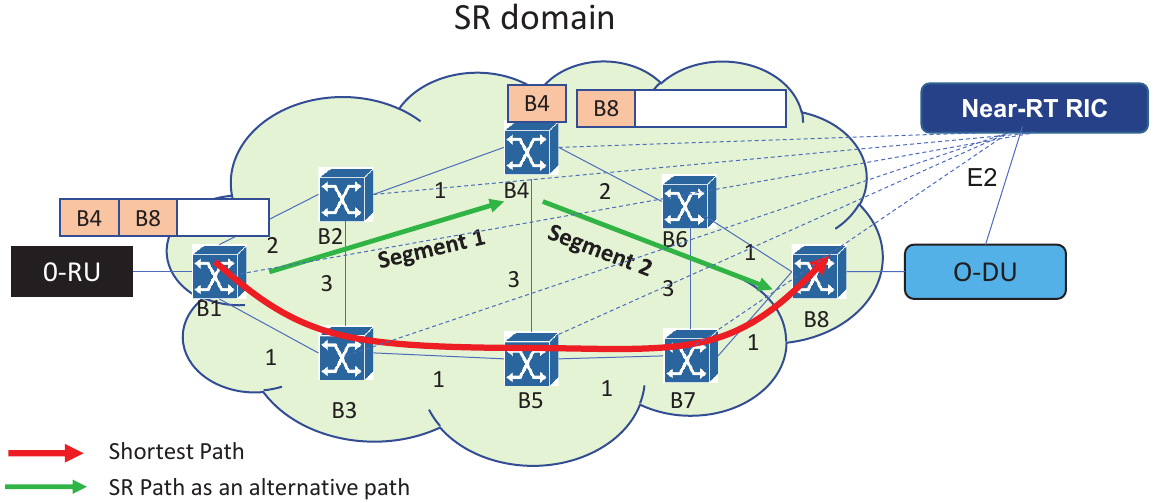}
	\caption{Illustration example of segment routing in FBN.}
	\label{fig:segment_routing}
\end{figure}

As an illustrative example, we consider SR in Fig. \ref{fig:segment_routing}, where the SR domain is defined as a set of TSNBs participating in the source-based routing model. In other words, O-RUs and O-DUs are connected to the segment domain but not included in the SR domain. Since we use VXLAN, we can have multiple fronthaul segment domains over one physical FBN, where each Near-RT RIC controls one domain. The ingress TSNB $B1$ adds the segment label $B4$ and $B8$ to the packet header, where  $B8$ is the destination address. The packet is routed from $B1$ to $B4$ along the shortest path $B1-B2-B4$. The top label is popped at bridge $B4$, and the packet is routed to $B8$. In other words, we have two segments, $B1-B2-B4$ and $B4-B6-B8$. Routing within each segment is done by the  Interior Gateway Protocol (IGP) routing protocol such as Open Shortest Path First (OSPF). In other words, each link is associated with cost, and IGP can use the cost to choose the shortest path. The path (red path) from source to destination can be the shortest, but it does not guarantee to be the fastest route. Also, the shortest path may fail. Therefore, we consider the SR path (green path) as the best way to choose an alternative path based on the network delay.
\subsubsection{Segment Routing in Fronthaul Network}
We assume that ECs are close to the edge devices. Therefore, we use two segments in SR for FBN. As highlighted in \cite{hao2016optimizing}, SR with two segments is generally enough in traffic planning problems that aim to route traffic so that no link is overloaded. However, extending  from two segments to multiple segments is straightforward. 

In our SR, Near-RT RIC  chooses intermediate TSNB $w \in \mathcal{W}$ that splits the fronthaul path into two segments. Let us consider $i  \in \mathcal{W}$ as source TSNB and $j  \in \mathcal{W}$ as destination  TSNB.    When $w \neq i$ and $w \neq j$, the offloaded traffic is first routed on the shortest path   $\Psi_{i,w}$ from $i$ to $w$ and then on the shortest path   $\Psi_{w,j}$ from $w$ to $j$.  To route the offloaded traffic via  FBN, the amount of traffic from $i$ to $j$ through intermediate TSNB $w$ is given by:
\begin{equation}
	l^w_{i,j}= \sum \nolimits_{\mathcal{V}^w_{i,j}\subset \mathcal{V}}x_v^{m\rightarrow n}d_v,
	\label{eq:interalloadFronthaul}
\end{equation}
where $\mathcal{V}^w_{i,j}\subset \mathcal{V}$ is a set of edge devices using fronthaul path $\Psi^w_{i,j}=\Psi_{i,w} \cup \Psi_{w,j}$  from $i$ to $j$ through intermediate TSNB $w$. The  important node $w$ as middle point for traffic routing can be chosen based on node centrality approach presented in \cite{trimponias2017centrality}.  
We use $\tau_{SR}^{v,m \rightarrow n}$ as  transmission delay of each fronthaul link  using SR, where $\tau_{SR}^{v,m \rightarrow n}$ is given by:
\begin{equation}
	\tau_{SR}^{v,m \rightarrow n}=\frac{ l^w_{i,j}}{\omega^j_i}. 
\end{equation}
However, when the fronthaul network is not segmented, the offloaded traffic is routed using a shortest path $\Psi_{i,j}$.  Transmission delay $\tau_{SP}^{v,m \rightarrow n}$ for shortest path becomes $\tau_{SP}^{v,m \rightarrow n}=\frac{l_{i,j}}{\omega^j_i}$, where $l_{i,j}=\sum \nolimits_{\mathcal{V}_{i,j}\subset \mathcal{V}}x_v^{m\rightarrow n}d_v$. Here,  $\mathcal{V}_{i,j}\subset \mathcal{V}$ is a set of edge devices using fronthaul shortest path $\Psi_{i,j}$  from $i$ to $j$.

Considering the shortest path  $\Psi_{i,j}$ and the SR path  $ \Psi^w_{i,j}$, the Near-RT RIC needs to choose one path  that gives the lowest possible latency to reach egress node. Therefore, we define fronthaul path selection variables, where $\eta_{SP}^{m \rightarrow n}$ is for SP and  $\eta_{SR}^{m \rightarrow n}$ is for the the SR, such that $\eta_{SP}^{m \rightarrow n}+ \eta_{SR}^{m \rightarrow n}=1$. 
\begin{equation}
	\setlength{\jot}{10pt}
	\eta_{SP}^{m \rightarrow n} =
	\begin{cases}
		1,\; \text {if 
			 $\tau_{SP}^{v,m \rightarrow n} \leq \tau_{SR}^{v,m \rightarrow n}$ }, \\
		0,\;\text{otherwise,}
	\end{cases}
\end{equation}
\begin{equation}
	\setlength{\jot}{10pt}
\eta_{SR}^{m \rightarrow n} =1- \eta_{SP}^{m \rightarrow n}
\end{equation}
If the shortest route has more transmission delay than the SR path, the ingress TSNB uses the SR path.  Therefore, the fronthaul transmission delay becomes:
\begin{equation}
	\tau_v^{m \rightarrow n}=\eta_{SP}^{m \rightarrow n} \tau_{SP}^{v,m \rightarrow n}+ \tau_{SP}^{v,m \rightarrow n}\eta_{SR}^{m \rightarrow n}.
\end{equation}

\subsection{Computation at Edge and Regional Clouds}
\label{subsec:massiveData_handlingNode} 
\subsubsection{Computation at Edge Clouds}
\label{subsec:EdgeComputation}

Once tasks reaches O-DU, the O-DU sends it to MEC server $r$ via O-CU-UP and UPF.  Then, MEC server checks if it has computation resource $\chi_{vn}$ required to compute task $\Gamma_{v}$ from  edge device $v$. The $\chi_{vn}$ can be computed as follows:
\begin{equation}
\label{eq:compution_resource_allocation}	
	\chi_{vn}=\chi_n\frac{\tilde{z}_{v}}{\sum_{g \in \mathcal{V}_n}\tilde{z}_{g}},\; \forall v  \in \mathcal{V}_n,\; n \in \mathcal{N},
\end{equation}
where $\mathcal{V}_n$ is a set of edge devices connected to EC $n$. In Eq. (\ref{eq:compution_resource_allocation}), we use weighted proportional allocation, which is available in systems such as 4G and 5G cellular networks for resource allocation \cite{tun2019weighted}.

If $\chi_n-\chi_{vn} \leq \Theta_n$, the EC $n$ allocates $\chi_{vn}$ to the  task $\Gamma_{v}$.  Here, $\Theta_n$ is resource allocation threshold of EC $n$.
Furthermore, we define $y_v^{m \rightarrow n}$ as a computation decision variable, where $y_v^{m \rightarrow n}$ is given by:
\begin{equation}
	\setlength{\jot}{10pt}
	y_v^{m \rightarrow n} =
	\begin{cases}
		1,\; \text{if $\Gamma_{v}$ offloaded via O-RU $m$}\\
		\; \; \; \;\text{is computed at  EC $n$ ($\chi_n-\chi_{vn} \leq \Theta_n$)},\\
		0, \;\text{otherwise.}
	\end{cases}
\end{equation}
The total computation resource allocations must satisfy:
\begin{equation}
	\sum_{v\in \mathcal{V}_n}x_v^{m \rightarrow n}\chi_{vn}y_v^{m \rightarrow n}\leq \chi_n,\; \forall  n \in \mathcal{N}.
\end{equation}.
Using the computation resource $\chi_{vn}$, the execution latency $	\tau_{vn} $ of task  $\Gamma_{v}$  from edge device $v$ at   EC $n$ becomes:
\begin{equation}
	\label{eq:compution_time_mec}
	\tau_{vn} =\frac{d_v \tilde{z}_{v}}{\chi_{vn}}.
\end{equation}
Furthermore, the total execution and offloading time for task $\Gamma_{v}$ at  EC $n$ is given by:
\begin{equation}
	\tau^e_{vn}= \tau^{v \rightarrow m}  + \tau_v^{m \rightarrow n} +	\tau_{vn}.
\end{equation}

When $\chi_n-\chi_{vn} > \Theta_n$ or $\tilde{z}_{v}>\tau^e_{vn}$ or $\tau_{vn}> \tilde{\tau}_{v}$, we consider EC $n$ to be overloaded. EC $n$ requests support to any neighboring EC $q$ that has enough resources to satisfy the offloading demand and is located in less distance than RC by redirecting a task with high computation deadline. Otherwise, EC $n$ requests support RC. As proposed in  \cite{ndikumana2019joint}, we assume that ECs exchange the resource utilization information. The  EC $n$ checks resource utilization information of neighboring ECs,  then compares EC $q$ that has enough resources with RC using propagation delay. We use 
$\tau^{n \rightarrow q}$  to denote propagation delay between EC $q$ and EC $n$, where  $\tau^{n \rightarrow q}$ can be expressed as follows: 
\begin{equation}
	\tau^{n \rightarrow q}=\frac{L^{n \rightarrow q}}{\kappa},\; \forall r,q \in \mathcal{R},
\end{equation}
where $L^{n \rightarrow q}$ is the length of physical link between EC $q$ and EC $n$ and $\kappa$ is the propagation speed. Furthermore, the propagation delay $\tau^{n \rightarrow RC}$ between EC $n$ and RC can be expressed as follows: 
\begin{equation}
	\tau^{n \rightarrow RC}=\frac{L^{n  \rightarrow DC}}{\kappa},\; \forall  n \in \mathcal{N},
\end{equation}
where $L^{n\rightarrow {RC}}$ is the length of physical link between EC $n$ and RC. We define a task forwarding decision variable $x^{n \rightarrow q}_v$, which indicates whether or not the task of edge device $v$ is forwarded from  EC $n$ to   EC $q$ for computation. $x^{n \rightarrow q}_v$  is given by:
\begin{equation}
	\setlength{\jot}{10pt}
	x^{n \rightarrow q}_v =
	\begin{cases}
		1,\; \text{if $\tau^{n \rightarrow q} \leq \tau^{n \rightarrow RC}$,}\\
		0,\;\text{otherwise.}
	\end{cases}
\end{equation}

The execution latency $\tau_{vq}$  of task  $\Gamma_{v}$ at   EC $q$ can be calculated using a similar approach as in (\ref{eq:compution_time_mec}). Therefore, the total execution time for a  task offloaded by edge device $v$ to  EC $q$ becomes:
\begin{equation}
	\tau^e_{vnq}=\tau^{v \rightarrow m} +\tau_v^{m \rightarrow n} + \tau^{n \rightarrow q}_v+ \tau^{n \rightarrow q} + \tau_{vq}.
\end{equation}
Furthermore, the offloading delay $\tau^{n \rightarrow q}_v$ between  EC $n$ and  EC $q$ can be calculated as follows:
\begin{equation}
	\tau^{n \rightarrow q}_v=\frac{\sum_{v \in \mathcal{V}_n}x^{n \rightarrow q}_v d_v}{\omega^q_n}, \;\forall r,\;q \in \mathcal{R},
\end{equation}
where $\omega^q_n$ is link capacity between EC $n$ and  EC $q$.

\subsubsection{Offloading Tasks to the Regional Cloud}
\label{subsubsec:massiveData_handlingDC} 

When there are no available computation resources at any neighboring EC $q$, or  EC $q$  is at far distance than RC,  the EC $n$ forwards the task to the RC. Therefore, we define $x^{n \rightarrow RC}_v$ as offloading decision variable that indicates whether or not the task of edge device $v$ is offloaded  by EC $n$ to  the RC:
\begin{equation}
	\setlength{\jot}{10pt}
	x^{n \rightarrow RC}_v=
	\begin{cases}
		1,\; \text{if $\tau^{n \rightarrow q} > \tau^{n \rightarrow RC}$ or no available resources}\\
		\; \; \; \;\text{at ECs},\\
		0, \;\text{otherwise.}
	\end{cases}
\end{equation}
We define $\tau^{n \rightarrow RC}_v$ as the offloading delay between EC $n$ and RC, where $\tau^{n \rightarrow RC}_v$ is given by:
\begin{equation}
	\tau^{n \rightarrow RC}_v=\frac{\sum_{v \in \mathcal{V}_n}x^{n \rightarrow RC}_v d_v}{\omega^{RC}_n}, \;\forall n,\; q \in \mathcal{N}.
\end{equation}
$\omega^{RC}_n$ is the link capacity between  EC $n$ and RC. Therefore, the total execution time for task $\Gamma_{v}$  offloaded by edge device $v$ at RC becomes:
\begin{equation}
	\tau^e_{vnRC}= \tau^{v \rightarrow m}  + \tau_v^{m \rightarrow n} + \tau^{n \rightarrow RC}_v  + \tau^{n \rightarrow RC} + \tau_{vRC},\;
\end{equation}
where $\tau_{vRC}$ can be calculated using a similar approach as in (\ref{eq:compution_time_mec}). 

The total offloading and computation latency $\tau^\textrm{off}_{v}$ of task $\Gamma_{v}$  from edge device $v$ is given by:
\begin{equation}
	\label{eq:compution_off}
	\begin{aligned}
		\tau^\textrm{off}_{v}=y_v^{m \rightarrow n}\tau^e_{vn} + (1-y_v^{m \rightarrow n}) (x^{n \rightarrow q}_v\tau^e_{vnq}+ x^{n \rightarrow RC}_v\tau^e_{vnRC}).
	\end{aligned} 
\end{equation}
To ensure that task $\Gamma_{v}$ from edge device $v$ is executed at only one location, i.e., computed locally at a edge device, or at one of ECs, or at RC, we impose the following constraints:
\begin{equation}
\begin{aligned}
	\label{eq:compution_off_contraint}
	\setlength{\jot}{10pt}
	(1-x_v^{m \rightarrow n}) + x_v^{m \rightarrow n}(y_v^{m \rightarrow n}+ n_v^\textrm{sup})= 1,
\end{aligned}
\end{equation}
where $n_v^\textrm{sup}=1-y_v^{m \rightarrow n}(x^{n \rightarrow q}_v + x^{n \rightarrow RC}_v)$ corresponds to the support EC $n$ gets from neighboring EC $q$ or RC to compute the offloaded task $\Gamma_{v}$.
\section{Problem Formulation and Solution} 
\label{sec:Problem_Formulation_Solution}
This section discusses the problem formulation for minimizing total delay, including offloading, fronthaul routing, and computation delays. Then, we present the solution approach of the formulated problem.
\subsection{Problem Formulation} 
\label{subsec:Problem_Formulation}
Computing task  $\Gamma_{v}$ locally at the edge device   $v$  requires computational delay cost $\tau^\textrm{loc}_v$.  On the other hand, computing offloaded task  $\Gamma_{v}$ at the EC or RC requires offloading, fronthaul routing, and cloud computation delays $\tau^\textrm{off}_v$. We assume that the offloading decision making operates in  time frames $t \in  \mathcal{T}= \{1,2, \dots ,  T\}$. Therefore, considering both local computation and offloading at time $t$, we formulate the following optimization problem to minimize total delay.
\begin{subequations}
	\label{eq:problem_formulation0}
	\begin{align}
		&\underset{\vect{x},\vect{\eta}, \vect{y} }{\text{min}}\ \  \sum_{n\in \mathcal{N}}\sum_{v\in \mathcal{V}_n} (1-x_v^{m \rightarrow n}(t))\tau^\textrm{loc}_{v}(t) +x_v^{m \rightarrow n}(t)\tau^\textrm{off}_{v}(t)
		\tag{\ref{eq:problem_formulation0}}\\
		& \text{subject to}\nonumber\\
		& \sum_{v\in \mathcal{V}_n}x_v^{m \rightarrow n}(t)b_v^m (t)\leq 1, \;  \forall m  \in \mathcal{M},\label{first:a}\\
		& x_v^{m \rightarrow n}(t)(\,\eta_{SP}^{m \rightarrow n}(t) l_{i,j} (t)+ \eta_{SR}^{m \rightarrow n}(t)l^w_{i,j}(t))\, \leq\omega^j_i(t),\label{first:b}\\
		&\sum_{v\in \mathcal{V}_n}x_v^{m \rightarrow n}(t)\chi_{vn}(t)y_v^{m \rightarrow n}(t)\leq \chi_n(t),\; \forall  n \in \mathcal{N}. \label{first:c}
	\end{align}
\end{subequations}
The objective function in  (\ref{eq:problem_formulation0}) combines (\ref{eq:local}), (\ref{eq:compution_off}), and (\ref{eq:compution_off_contraint}). The constraint in (\ref{first:a}) guarantees that the sum of wireless resources allocated to all edge devices has to be less than or equal to the total available resources. The constraint in (\ref{first:b}) is related to FBN, ensuring that each TSNB does not send more traffic than the link capacity. The constraint in (\ref{first:c}) guarantees that the computation resources allocated to edge devices at each EC do not exceed available computation resources.

The problem in (\ref{eq:problem_formulation0}) is combinatorial optimization problem, which is NP-hard.  Also, using combinatorial optimization, the number of possibilities increases exponentially as the problem size increases. However, a heuristic approach can be designed to solve it. As an example, by applying Block Successive Majorization Minimization (BS-MM) technique described in \cite{sun2016majorization, ndikumana2020deep}, we  can get a proximal convex surrogate problem  by adding the quadratic term to (\ref{eq:problem_formulation0}) and relaxing variables. Then,  we can minimize the proximal convex surrogate problem. However, the heuristic approach may results in a stationary solution, which is not appropriate for a dynamic environment. Therefore, we change the problem in (\ref{eq:problem_formulation0})  to be a reward function $	r_t$ which can be maximized by an existing ML approach such Deep Reinforcement Learning (DRL) \cite{sutton2018reinforcement}. Also, $r_t$  can represent network condition:
		\begin{multline}
			\label{eq:problem_formulation1}
			r_t=  \varpi_\textrm{CoD}(\sum_{n\in \mathcal{N}}\sum_{v\in \mathcal{V}_n} \tilde{\tau}_{v}(t) - ((1-x_v^{m \rightarrow n}(t))\tau^\textrm{loc}_{v}(t)  +\\ x_v^{m \rightarrow n}(t)\tau^\textrm{off}_{v}(t))\, +	 \varpi_w(1-\sum_{v\in \mathcal{V}_n}x_v^{m \rightarrow n}(t)b_v^m (t)) + \\ \varpi_f (\omega^j_i(t)-x_v^{m \rightarrow n}(t)(\,\eta_{SP}^{m \rightarrow n}(t) l_{i,j} (t)+ \eta_{SR}^{m \rightarrow n}(t)l^w_{i,j}(t)) + \\ \varpi_c (\chi_n(t) -\sum_{v\in \mathcal{V}_n}x_v^{m \rightarrow n}(t)\chi_{vn}(t)y_v^{m \rightarrow n}(t)) .
	\end{multline}
	In the reward function (\ref{eq:problem_formulation1}),  $\varpi_\textrm{CoD}$ is the Cost of Delay (CoD), which represents the penalty for missing the computation deadline $\tilde{\tau}_{v}(t)$ in offloading. Similarly,  we use $\varpi_w$ to denote the penalty of violating wireless communication resource constraint, and $\varpi_f$ is the penalty for overloading the fronthaul link capacity. $\varpi_c$  is the penalty for violating computational resources constraint. The penalties $\varpi_\textrm{CoD}$, $\varpi_c$, $\varpi_f$, and $\varpi_c$ correspond to the constraints in (\ref{first:a}), (\ref{first:b}), and (\ref{first:c}), respectively. In other words, using penalties, $r_t$ goes down
(Figs $11$ and $12$)	when any of these constraints is violated.
\begin{figure}[t]
	\centering
	\includegraphics[width=1.0\columnwidth]{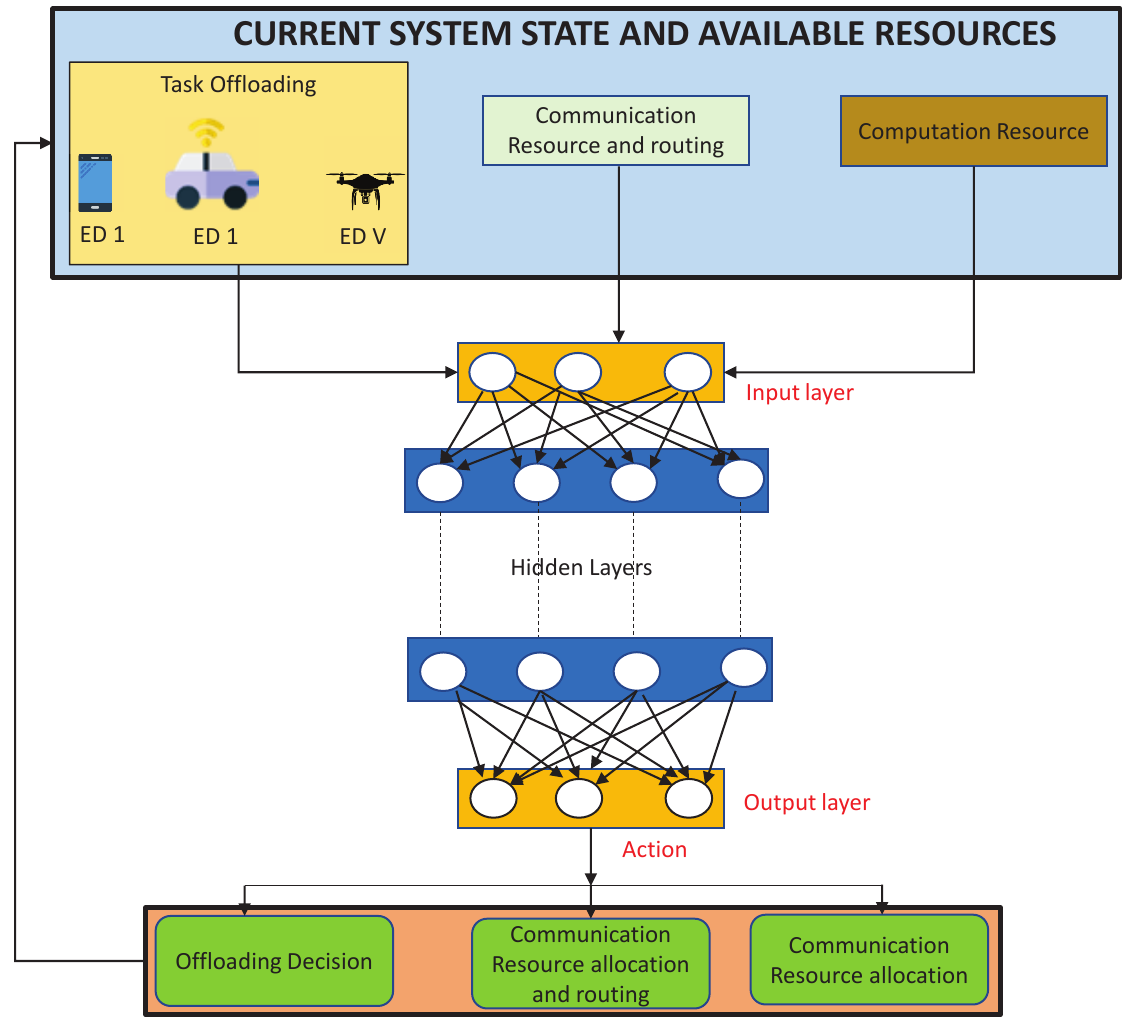}
	\caption{DRL process to solve (\ref{eq:problem_formulation1}).}
	\label{fig:RL}
\end{figure}
\subsection{Proposed Solution} 
\label{subsec:ProposedSolution}

One of the emerging approaches for handling (\ref{eq:problem_formulation1}) is to use ML algorithms such as DRL. In DRL, a DRL agent acts in the environment to find a proximal solution for a given problem. DRL considers time-varying workloads and network conditions, where the Markov Decision Process (MDP) can be applied to model the environment (detailed procedures of MDP are defined in \cite{feinberg2012handbook}). In our  solution approach presented in Fig. \ref{fig:RL} , we consider tuple $\langle \mathcal{A}, \;\mathcal{S}, \; \mathcal{R}, \; \mathcal{P}  \rangle$ defined as follows:
\begin{itemize}
	\item 
	 We consider the action space $ \mathcal{A} = {\{(\vect{x}, \vect{b}, \vect{\eta}, \vect{y}) \}}$ that  represents the offloading decisions, communication resource allocation and routing decisions, and computation resources allocation decision.
	\item 
	We define a state space $\mathcal{S}$ which  consists of current  local computation, offloading using wireless and fronthaul,  and  cloud computation states such that $\mathcal{S}= {\{(\vect{\tau}^\textrm{loc}, \vect{\tau}^\textrm{off}, \vect{\omega}, \vect{l}, \vect{\chi})\}}$.
	\item We represent $\mathcal{R}=\sum_{t\in \mathcal{T}}r_t$ as 
	an accumulated reward which  indicates how action chosen in particular state improves resource allocation and CoD;
	\item  We use  $\mathcal{P}= {\{(p(s_{t+1}, r_t|s_t, a_t)\}}$ as the transition probability matrix that governs transition dynamics from one state $s_t \in \mathcal{S}$ to another $s_{t+1} \in \mathcal{S}$ in response to action $a_t \in \mathcal{A}$ and reward $r_t \in \mathcal{R}$.
\end{itemize}

The state transition and reward are stochastic and modeled
as an MDP. The state transition probabilities and rewards depend only on the state of the offloading environment and the agent's action. The transition from $s_t \in \mathcal{S}$  to $s_{t+1} \in \mathcal{S}$ with reward $r_t \in \mathcal{R}$ when action $a_t \in \mathcal{A}$ chosen is characterized by the conditional transition probability $\mathcal{P}$,  which is only determined by offloading environment.

In Fig. \ref{fig:RL} (ED means edge device), the MEC server as an agent periodically learns to take actions, observes the most reward, and automatically adjusts its strategy. In DRL, we use Deep Q-Learning (DQL) \cite{fan2020theoretical}. We consider DQL as a better solution that leverages deep neural networks (DNNs) to train the deep learning model. In other words, DQL integrates deep learning into Q-Learning. The simplest form of  Q-Learning, which is called one-step Q-Learning, is given by:
\begin{equation}
	\begin{aligned}
	Q(s_t, a_t)= Q(s_t, a_t) + \alpha [r_{t+1} +\Upsilon Q(s_{t+1}, a) \\ - \; Q(s_{t}, a_t)],
	\end{aligned}
\end{equation}
where $\alpha$ is  the learning rate and $a \in \mathcal{A}$ is an action that was taken in the state 
$s_t$ by the agent. $\Upsilon$  ($0 <\Upsilon \leq 1$) is discount factor that encourages the agent to account more for short-term 
reward $r_t$. On the other hand, DQL uses standard feed-forward neural networks to calculate Q-Value. The DQL uses two networks,  Q-Network to  calculate Q-Value in the state $S_t$ and  target network to calculate Q-Value in the state $S_t+1$ such that:
\begin{equation}
	\begin{aligned}
		Q(s_t, a_t)= Q(s_t, a_t) + \alpha [r_{t+1} +\Upsilon\underset{\vect{a} }{\text{max}} Q(s_{t+1}, a) \\ - \; Q(s_{t}, a_t)].
	\end{aligned}
\end{equation}
\begin{figure}[t]
	\centering
	\includegraphics[width=1.0\columnwidth]{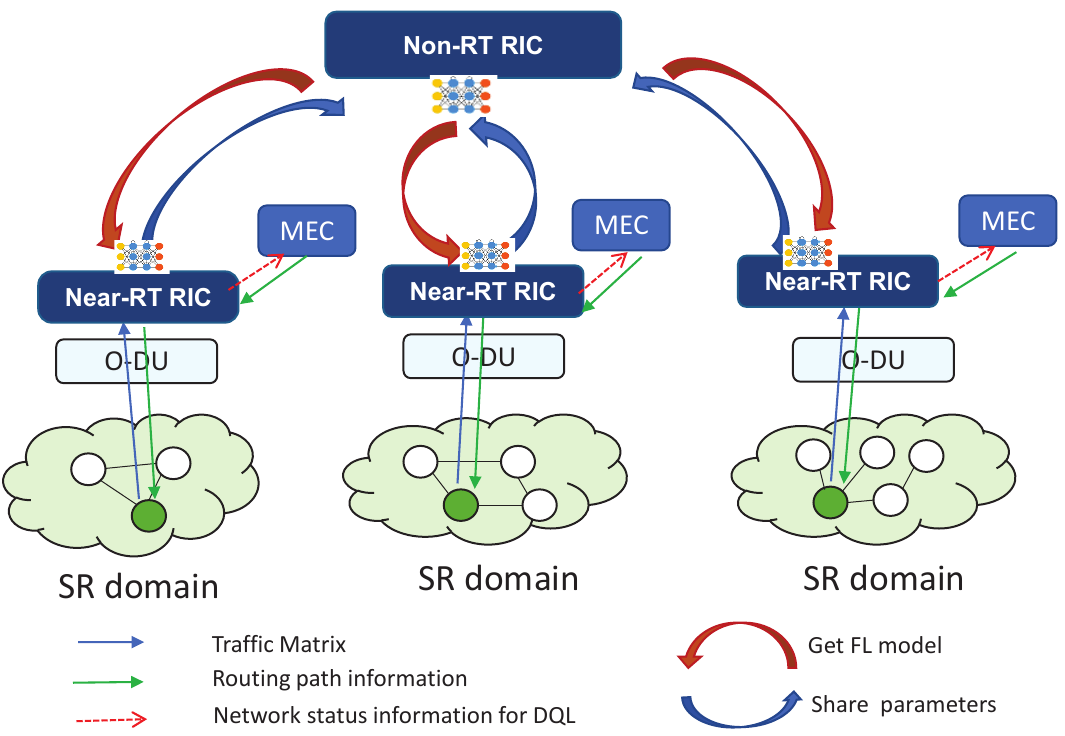}
	\caption{FL assisted DQL in routing.}
	\label{fig:Solution}
\end{figure}

As shown in Fig. \ref{fig:Solution},  Near-RT RIC has a global view of fronthaul resources and fronthaul routing in forms of the traffic matrix. Sending the whole traffic matrix to MEC for DQL can consume huge bandwidth. Near-RT RIC can use a network matrix to overcome this issue by predicting the appropriate candidate routes for fronthaul traffic. Then, Near-RT RIC sends processed fronthaul routing information to the MEC server for DQL. We consider processed fronthaul routing information to be smaller and take a short transmission time than sending the whole unprocessed traffic matrix. Since we have multiple Near-RT RICs,  we use Non-RT RICs to coordinate Near-RT RICs.  However,  sending traffic matrix to Non-RT RIC  for centralized fronthaul routing prediction may consume high bandwidth. To this end, we choose a Federated Learning (FL) approach over other approaches to learn fronthaul routing distributively. In FL, Near-RT RICs and Non-RT RIC are not required to exchange the whole traffic matrix but only the model and learning parameters. This can save bandwidth and guarantee the privacy of the fronthaul traffic matrix. 

 Each Near-RT RIC $n$ can get FL model and learning parameter $\vect{N}_t$ from Non-RT RIC. Then, Near-RT RIC $n$ uses traffic matrix  $\vect{l}_n^t$ of size $\varrho_n$ to improve the model through training and testing the downloaded model. We use $\vect{N}_t^1, \dots ,\vect{N}_t^{N}$ to denote the current parameters of the ECs. Each Near-RT RIC  $n \in \mathcal{N}$ computes its gradient $\vect{N}_t^n$, where $\vect{N}_t^n$ is given by:
\begin{equation}
	\begin{aligned}
		\vect{N}_t^n=\frac{1}{\varrho_n}\sum_{\varrho_n}^{}  
		\nabla f_{n,\varrho_n}(\vect{w_n},\vect{l}_n^t, \tilde{\vect{l}}_n^t ).
	\end{aligned}
\end{equation}
We use $f_{n,\varrho_n}$ as the loss function, $\vect{w_n}$ as the weight, and $\tilde{\vect{l}}_n^t$ as the predicted fronthaul routing at Near-RT RIC  $n$. Furthermore, Near-RT RIC $n$ calculates the difference $\vect{\Phi}_t^n$ between its gradients $\vect{N}_t^n $ and $\vect{N}_t$ as follows: 
\begin{equation}
	\vect{\Phi}_t^n=\vect{N}_t^n-\vect{N}_t, \;\forall n \in \mathcal{N}.
\end{equation}

Each Near-RT RIC $n$  shares $\vect{\Phi}_t^n$ with Non-RT RIC. Then,  Non-RT RIC aggregates the received parameters from Non-RT RICs. The parameter aggregation at Non-RT RIC is defined as follows:
\begin{equation}
	\vect{\varphi}_t=\frac{1}{N}\sum_{n=1}^{N}\vect{\Phi}_t^n
	\label{eq:parameter_aggragation_ec}.
\end{equation}
Furthermore, the  FL update at  Non-RT RIC can be expressed as follows:
\begin{equation}
	\vect{N}_{t+1}  =\vect{N}_t +	\alpha\vect{\varphi}_t.
	\label{eq:model_update_ec}
\end{equation}
Then, Non-RT RIC shares $\vect{N}_{t+1} $ with Near-RT RICs. The iteration continues until $\vect{N}_{t+1}  =\vect{N}_t$ (there is no more improvement of the FL model). We consider the FL model to be trained and tested once and saved for later usage to minimize the delay. Then,  Near-RT RIC  can load and use the pre-trained model.

A key  goal of FL is to optimize a global training objective function defined over distributed devices, where each device uses its data to optimize this global training objective \cite{haddadpour2019convergence}. In our approach, we use the loss function $f_{n,\varrho_n}$ as global training objective function, where each Near-RT RIC uses fronthaul traffic data to minimize  $f_{n,\varrho_n}$. As discussed and proved in \cite{haddadpour2019convergence},  local Stochastic Gradient Descent (SGD) with periodic averaging has $O(\frac{1}{\sqrt{NK}})$ convergence rate, where $N$ is the number of devices, and $K$ is the number of iterations. Therefore, our FL has $O(\frac{1}{\sqrt{NK}})$ convergence rate, where $N$ is the number of Near-RT RICs involved in FL.
\begin{algorithm}[t]
	\caption{: FL assisted DQL algorithm  for joint task offloading and segment routing.}
	\label{algo:DQL}
	\begin{algorithmic}[1]
		\STATE{\textbf{Input:} Get communication resources and routing (from Near-RT RIC using FL), computation, and offloading states at MEC server;}
		\STATE{\textbf{Output:}  Offloading variable $\vect{x}$, fronthaul rouring variable $\vect{\eta}$, computation variable $\vect{y}$, communication resources allocation  $\vect{b}$, and computation resources allocation $\vect{\chi}$;}
		\STATE{MEC server uses Target Network and Q-Network to get the Q-Values of all possible actions in the defined state;}
		\REPEAT
		\STATE{Pick random action $a_t$ or action $a_t$ with the maximum Q-Value from the set of actions $\mathcal{A}$  based on$\Upsilon$ value;}
		\STATE{Perform action $a_t$, observe reward $r_t$ and the next state $s_{t+1}$;}
		\STATE{Store $\langle s_{t}, s_{t+1}, \; a_t, \; 	r_t \rangle$ in the experience replay memory;}
		\STATE{Sample random batches from experience replay memory and perform training of the Q-Network;}
		\STATE{Each kth iteration, copy the weights values from the Q-Network to the Target Network;}
		\UNTIL{terminal state  is reached}
		\STATE{MEC server informs Near-RT RIC about updated communication resource allocation and fronthaul routing decision.}
	\end{algorithmic}
\end{algorithm}
 \begin{figure*}[t]
	\centering
	\includegraphics[width=2.0\columnwidth]{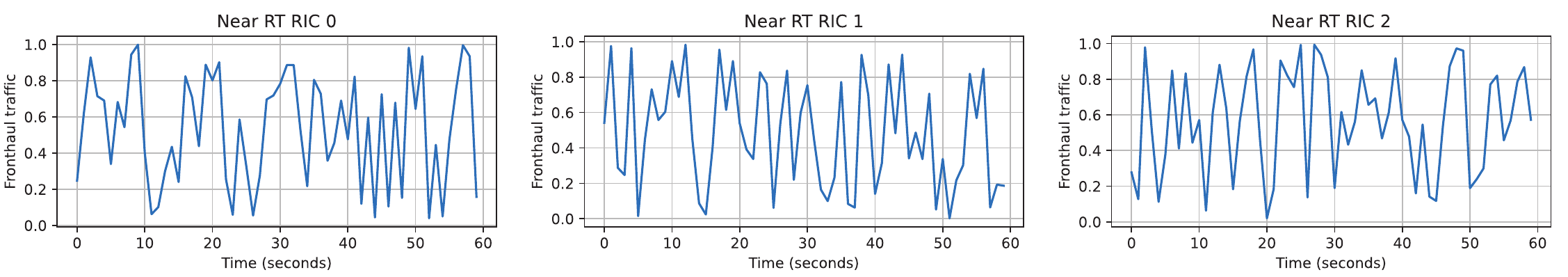}
	\caption{Normalized fronthaul traffic data at Near-RT RICs. }
	\label{fig:FL_Fronthaul}
\end{figure*}
\begin{figure}[t]
	\centering
	\begin{minipage}{0.5\textwidth}
		\centering	
		\includegraphics[width=1.0\columnwidth]{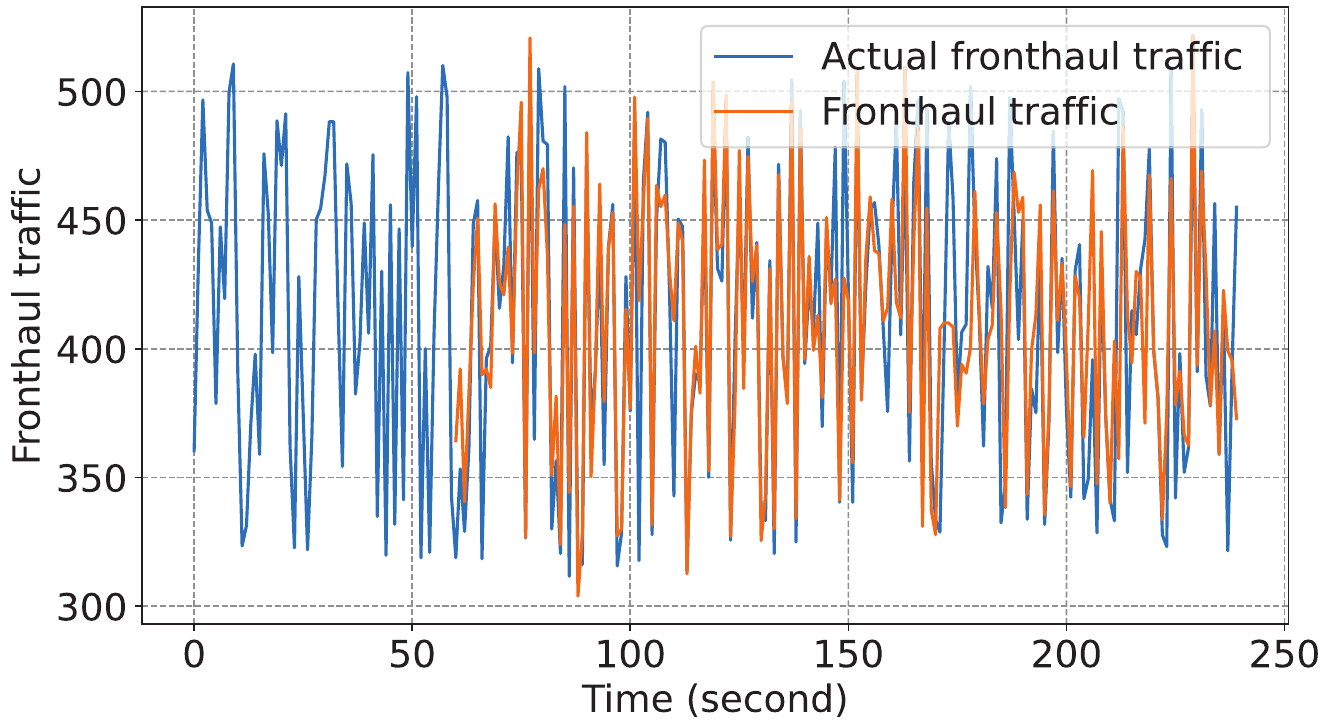}
		\caption{Prediction of fronthaul traffic (Mbps).}
		\label{fig:Fronthaul_Traffic}
	\end{minipage}
	\begin{minipage}{0.5\textwidth}
		\centering	
		\includegraphics[width=1.0\columnwidth]{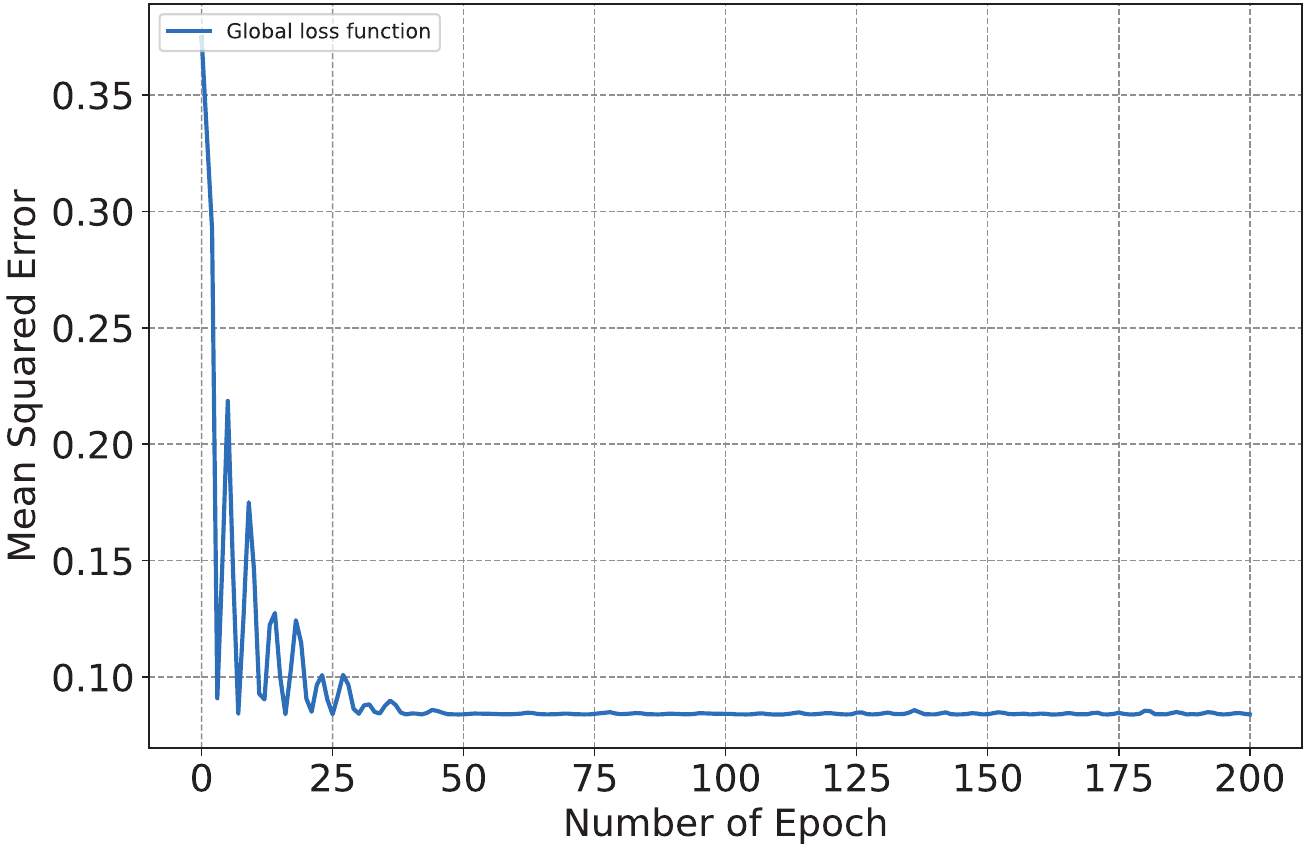}
		\caption{Convergence of global loss function $f_{n,\varrho_n}$.}
		\label{fig:loss}
	\end{minipage}
\end{figure}

For FL assisted DQL, we propose  Algorithm \ref{algo:DQL} to get offloading variable $\vect{x}$, fronthaul routing variable $\vect{\eta}$, computation variable $\vect{y}$, communication resources allocation  $\vect{b}$, and computation resources allocation $\vect{\chi}$.  In the Algorithm \ref{algo:DQL}, the MEC server gets communication resources and routing state from Near-RT RIC that uses FL, computation, and offloading states. Then, MEC server performs DQL processes. After DQL processes, the MEC server sends to Near-RT RIC  updated fronthaul routing decisions. In Fig.  \ref{fig:Solution},  after receiving fronthaul routing decisions, Near-RT RIC sends segment paths information to the ingress TSNB to push the segment labels on the header of incoming packets. Once the segments are added to the packet header,  fronthaul packets are routed through FBN using these segments.

\section{Performance Evaluation}
 \label{sec: SimulationResultsAnalysis}
 This section presents the performance evaluation of the proposed FL-assisted DQL for joint task offloading and fronthaul SR. 
 \begin{figure}[t]
	\centering
 	\begin{minipage}{0.5\textwidth}
 		\centering	
 		\includegraphics[width=1.0\columnwidth]{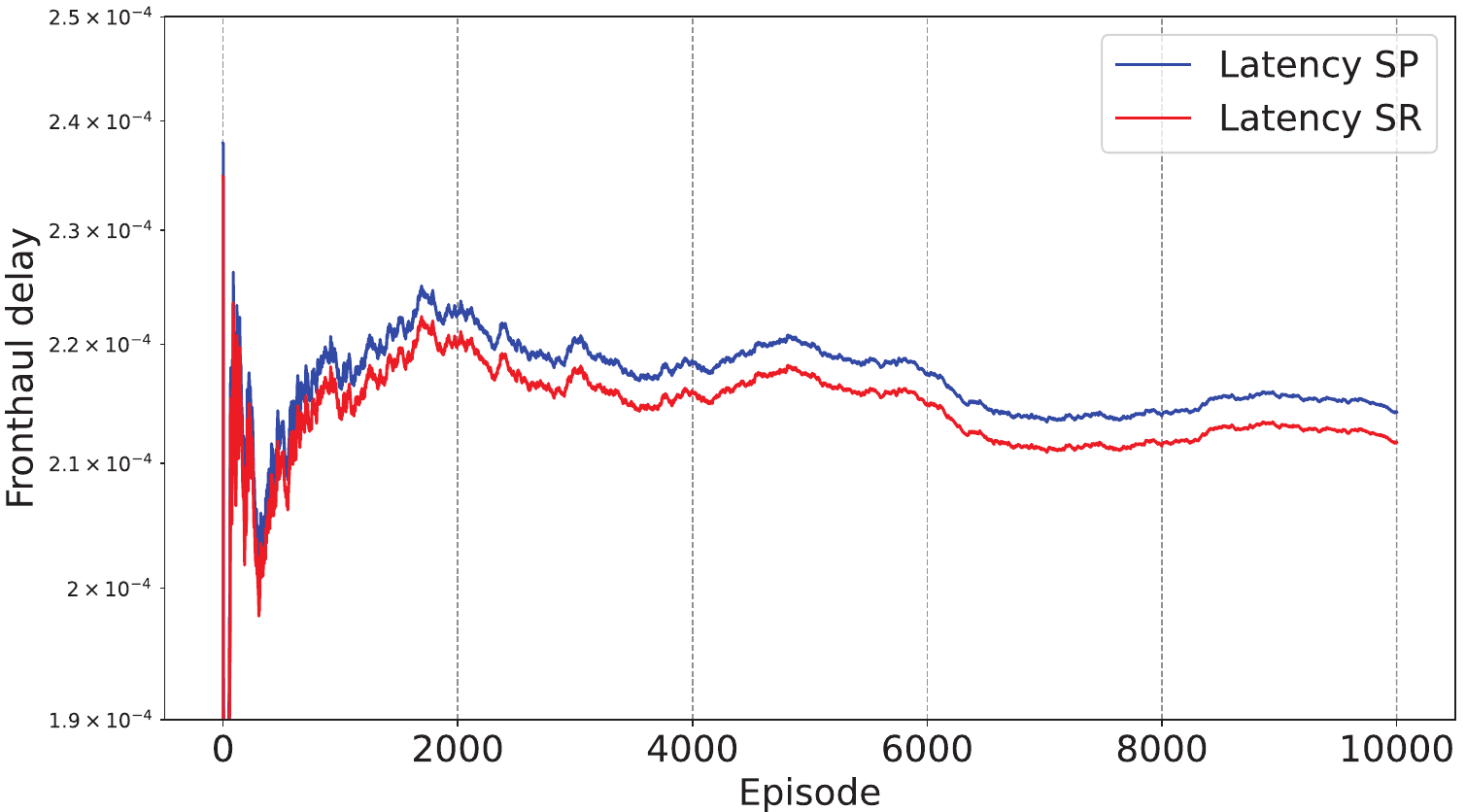}
 		\caption{Delay for shortest paths vs. SR paths.}
 		\label{fig:Fronthaul_Routing}
 	\end{minipage}
 \begin{minipage}{0.5\textwidth}
 	\centering	
 	\includegraphics[width=1.0\columnwidth]{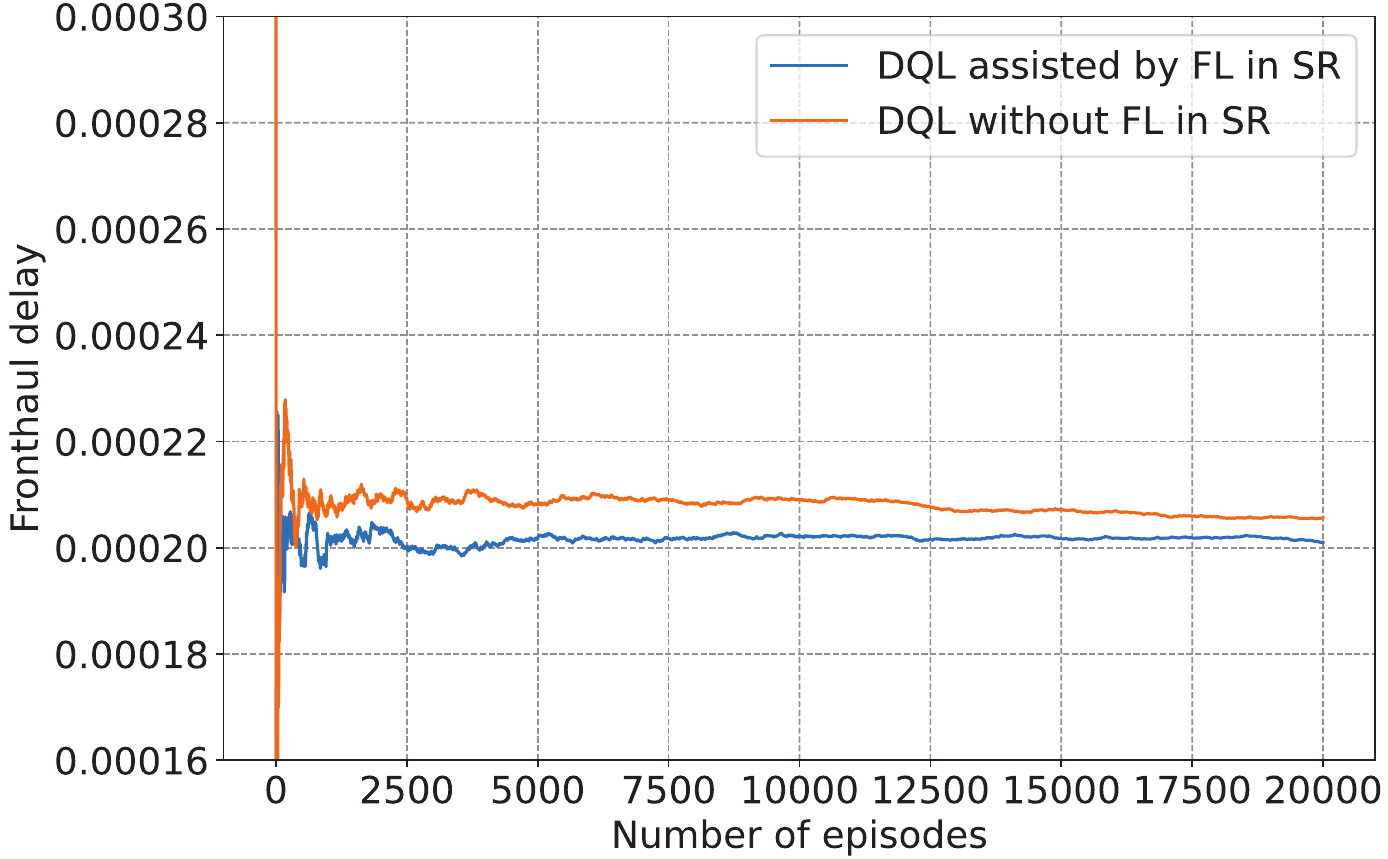}
 	\caption{DQL with and without FL assistance in SR.}
 	\label{fig:fronthaul_delayDQL}
 \end{minipage}
\end{figure}
\begin{figure}[t]
	\centering
 	\begin{minipage}{0.5\textwidth}
 	\centering	
 	\includegraphics[width=1.0\columnwidth]{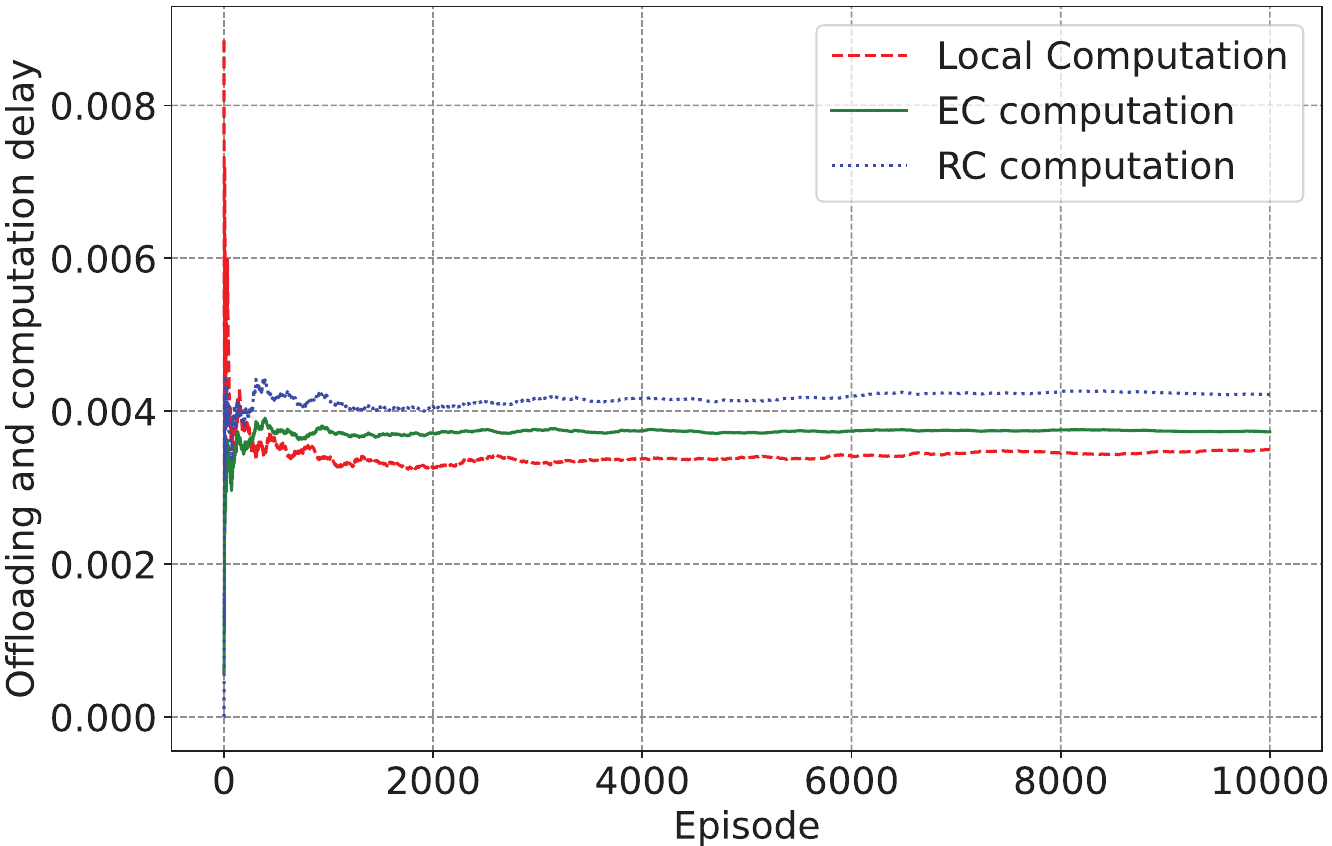}
 	\caption{Offloading and computation delay.}
 	\label{fig:OfflaodingComputation}
 \end{minipage}
	\begin{minipage}{0.5\textwidth}
	\centering	
	\includegraphics[width=1.0\columnwidth]{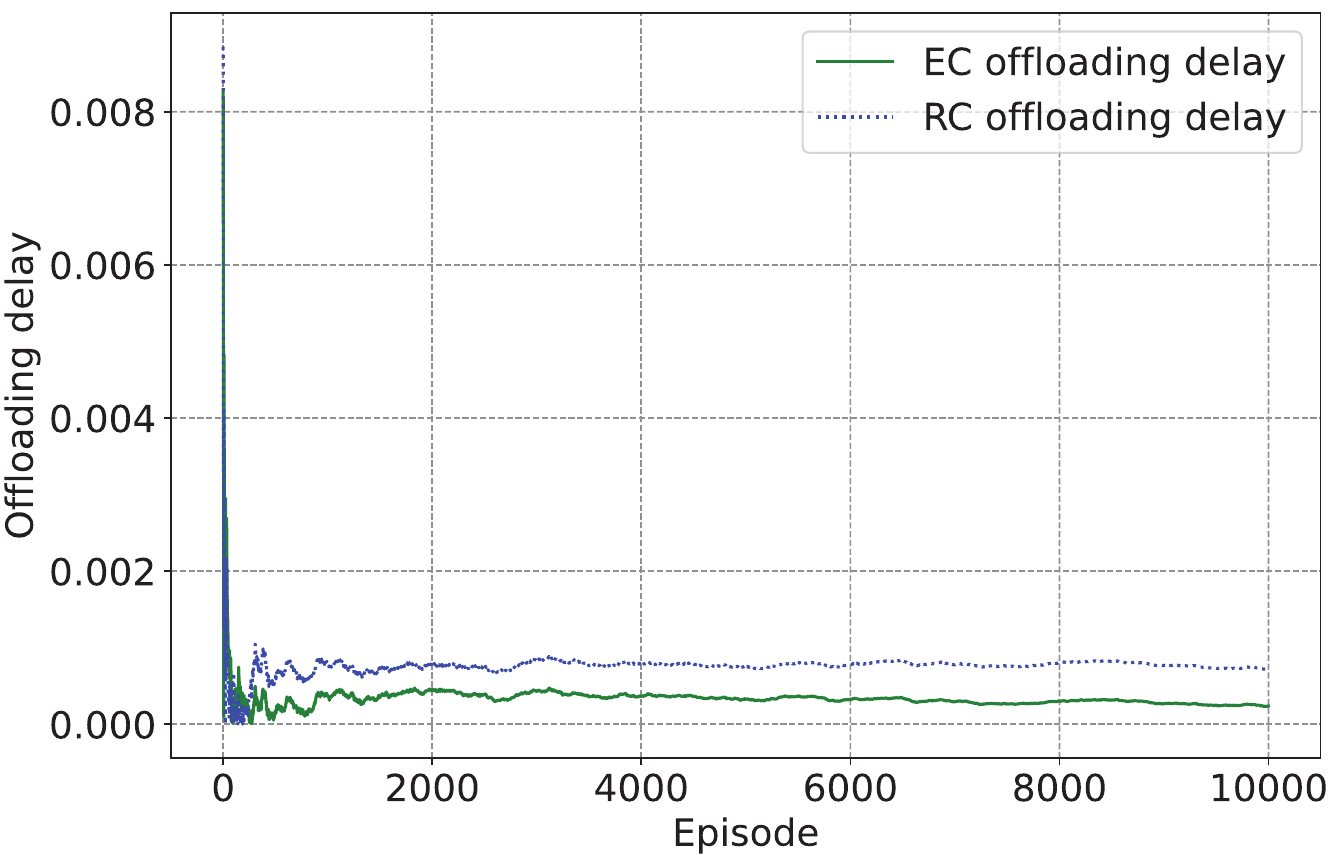}
	\caption{EC offloading vs RC offloading.}
	\label{fig:OfflaodingDelay}
\end{minipage}
 \end{figure}

\subsection{Simulation Setup}
In simulation setup, we use edge devices $V=65$, where $\tilde{z}_{v}= 737$ cycles per bit. For task $\Gamma_{v}$ of the edge device, the size of the input data ${d_v}$ is generated randomly within a range of $1$ to $8$ $Mb$. The task computation deadline for each device $v$ is  within a range of $\tilde{\tau}_{v}=0.2$ to $\tilde{\tau}_{v}=1.2$ seconds.  Furthermore, each edge device has a computation resource $\chi_v=2$ $GHz$.
To offload the computation task in the wireless network, we set  the transmission power $\rho_v=27.0$ dBm. The channel bandwidth is in the range from  $b_v^m=25$ MHz to $b_v^m=32$ MHz. For fronthaul routing, we use the cubical graph from  NetworkX (a Python library for studying graphs and networks) \cite{NetworkX} of 8 nodes. Each edge in graph has  bandwidth in range  $ \omega^j_i = 6000$ to $ \omega^j_i= 6500$ $Mbps$. The cubical graph is connected to three O-RUs and three ECs with links of bandwidth selected in the range from $6000$ to $6500$ $Mbps$. Since each EC has  O-DU and Near-RT RIC,  we consider each Near-RT RIC manages one virtual fronthaul segment routing domain as overlay networks that sit on top of one fronthaul physical network. Furthermore, we consider bandwidth between each pair of ECs that hosts O-DUs  in the range from $\omega^q_n=7000$ to $\omega^q_n =7500$ $Mbps$.  Also, the symmetric bandwidth between each EC and RC is selected in the range from  $\omega^{RC}_n = 7000$ to $\omega^{RC}_n=7500$ $Mbps$.
Each EC  $n$ has computation resource in the range from  $\chi_n=10$ $GHz$ to $\chi_n=30$ $GHz$, while at RC, the computation resource is in the range $\chi_{RC}=20$ to $\chi_{RC}=40$ $GHz$. 

\begin{figure}[t]
	\centering
	\begin{minipage}{0.5\textwidth}
		\centering
		\includegraphics[width=1.0\columnwidth]{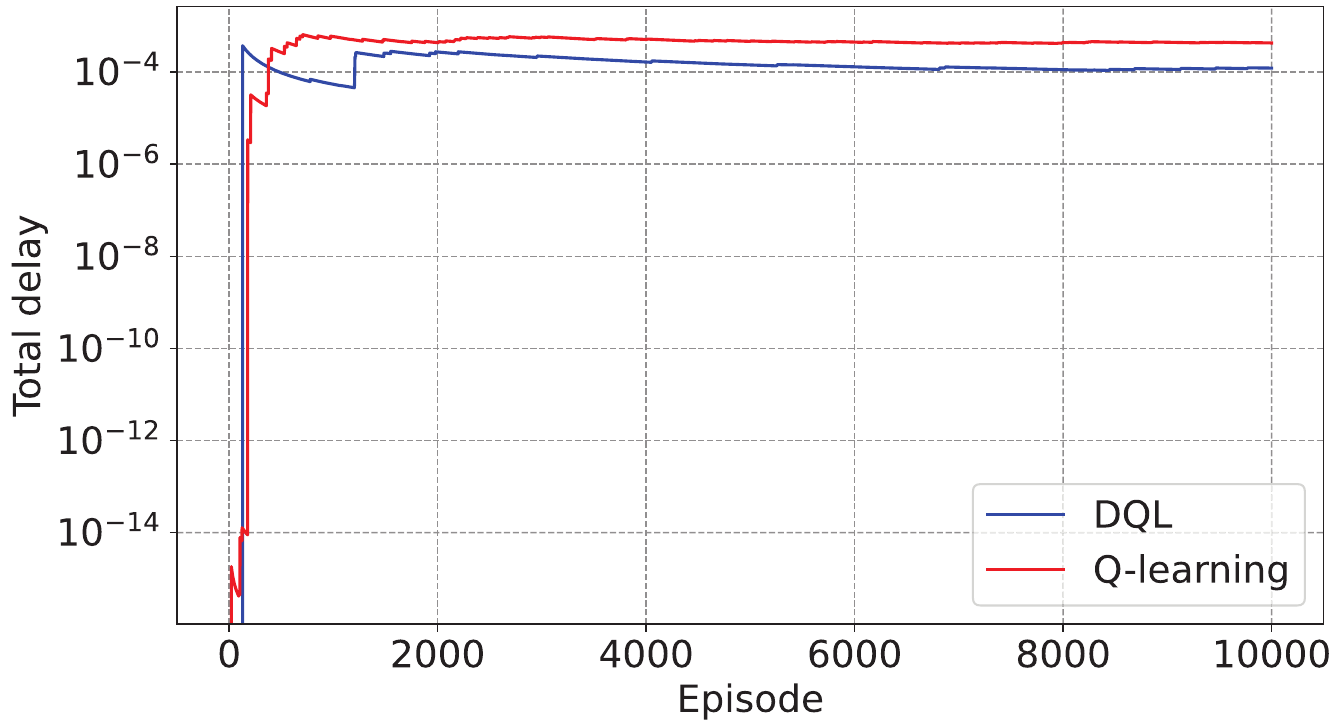}
		\caption{Total delay of DQL vs Q-Learning.}
		\label{fig:Running_Time}
	\end{minipage}
\begin{minipage}{0.5\textwidth}
\centering
\includegraphics[width=1.0\columnwidth]{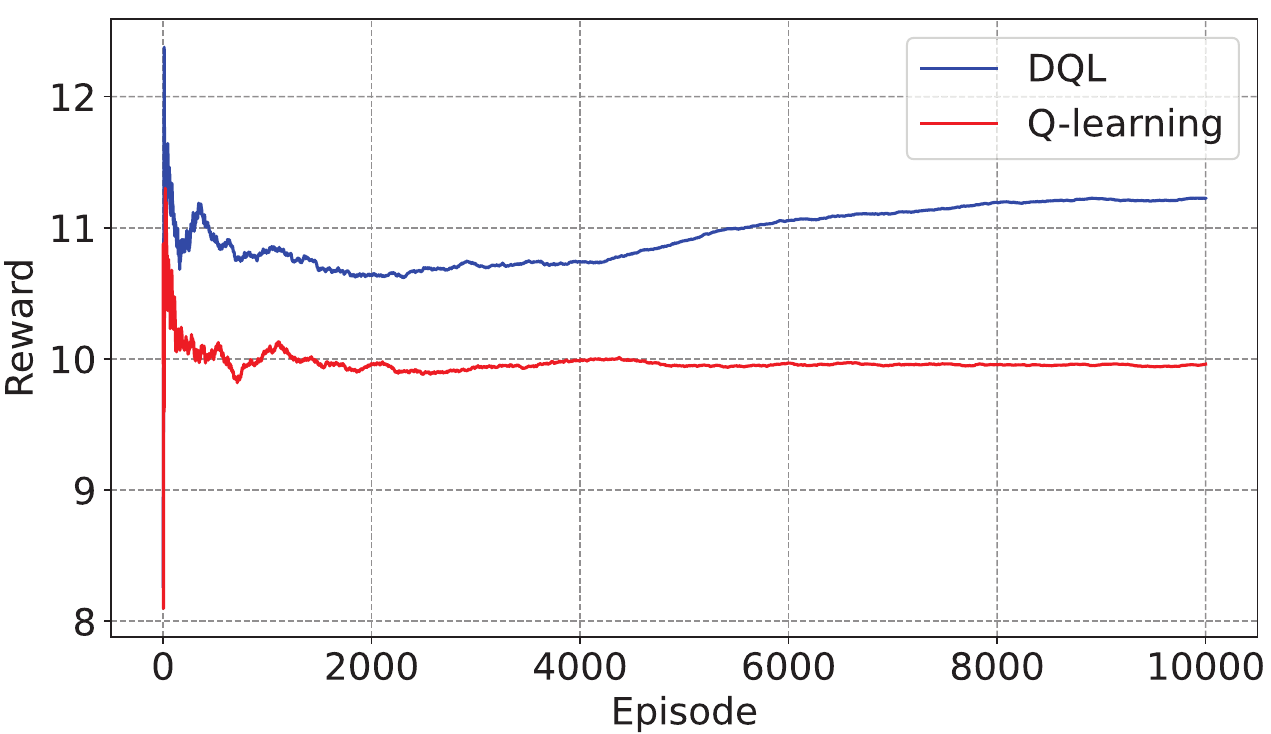}
\caption{Maximization of reward.}
\label{fig:Reward1}
\end{minipage}
\end{figure}
\begin{figure}[t]
	\centering
	\begin{minipage}{0.5\textwidth}
		\centering	
		\includegraphics[width=1.0\columnwidth]{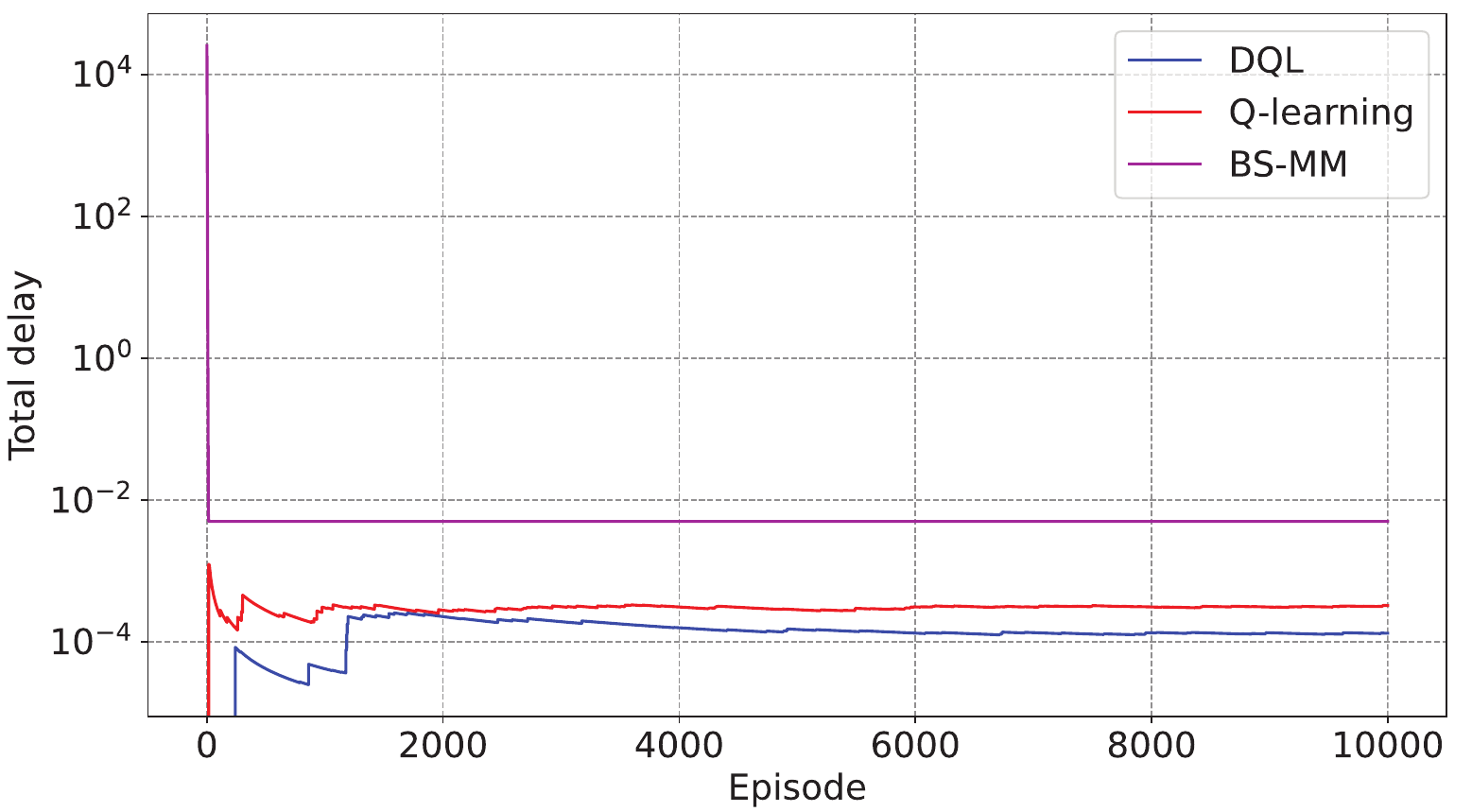}
		\caption{Total delay of BS-MM vs. DRL.}
		\label{fig:BS_MM_DRL}
	\end{minipage}
	\begin{minipage}{0.5\textwidth}
		\centering
		\includegraphics[width=1.0\columnwidth]{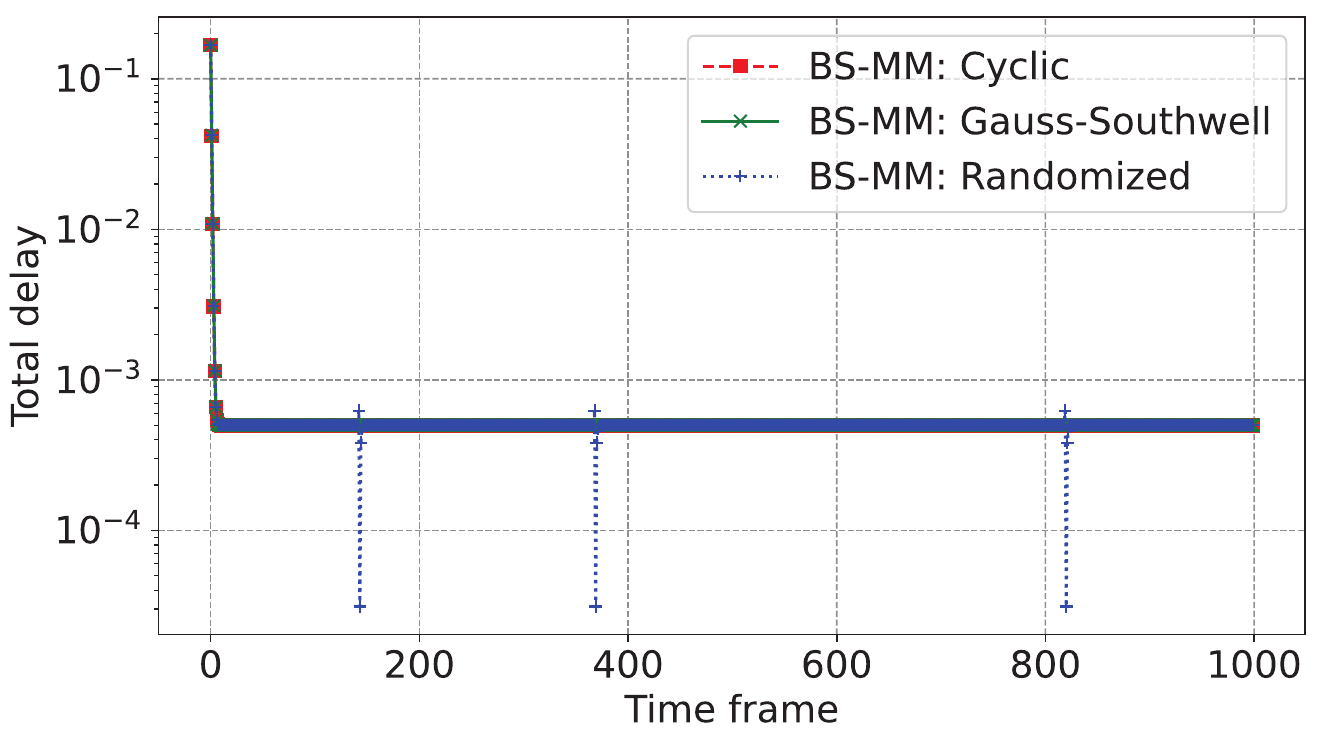}
		\caption{Delay for BS-MM in short time frame.}
		\label{fig:Total_delay_BS_MM}
	\end{minipage}
\end{figure}

 We use PyTorch \cite{paszke2019pytorch} and Gym \cite{brockman2016openai} as machine learning libraries to make Q-Network and target network for DQL. We set $\alpha = 0.001$ and  $\Upsilon=0.995$ for Q-Network and target network of $3$ fully connected layers. For FL, we use TensorFlow Federated \cite{TensorFlow} and Long short-term memory (LSTM) \cite{alom2018history} of two layers ($64$ neurons in the input layer and one neuron in output later) to predict fronthaul traffic in each fronthaul path. We generated a fronthaul traffic matrix using both shortest path and SR paths for  $3600$ seconds for FL. Based on the traffic matrix at Near-RT RICs,  Fig. \ref{fig:FL_Fronthaul} shows the normalized fronthaul traffic at each Near-RT RIC in the range between $0$ and $1$. The FL model was trained offline and saved in memory, where Near-RT RIC loads the pre-trained model for predicting the fronthaul routing. 
 
\subsection{Baseline Approaches}
In the performance evaluation, for fronthaul routing, we use Dijkstra's shortest path algorithm \cite{wei2021shortest} as a baseline for SP to compare with SR. SP and SR approaches use the cubical graph from  NetworkX \cite{NetworkX}. SP approach uses an unsegmented cubical graph and computes the shortest fronthaul path in the graph between source and destination. On the other hand, SR divides the cubical graph into segments, and this requires the computation of intermediate node $w \in \mathcal{W}$ that splits the fronthaul path into segments. Furthermore, for our joint task offloading, fronthaul segment routing, and edge computation problem, we consider  Q-Learning and BS-MM-based solutions described in Section \ref{sec:Problem_Formulation_Solution} as baselines. Then, we compare  DQL-based solution with Q-Learning and BS-MM-based solutions.
\subsection{Simulation Results}
Fig. \ref{fig:Fronthaul_Traffic} shows the sample of predicted fronthaul traffic at Near-RT RIC in terms of $Mbps$  using different paths. This figure starts having predicted fronthaul traffic after $60$ seconds because we use $60$ seconds as a lookback period. The lookback period defines the number of time steps used to predict fronhaul traffic. In other words, using our prediction approach, we can know fronthaul traffic in $60$  seconds ahead. In the initial implementation of our  fronthaul routing approach for a real network environment, the Near-RT RIC should record fronthaul traffic for at least a lookback period. During this period, Near-RT RIC can use existing routing approaches such as Open Shortest Path First (OSPF). Then, after the lookback period, the Near-RT RIC can start using SR and select the path between SP and SR with the lowest latency to reach each egress node. Furthermore, Fig. 8 presents the convergence of MSE as global loss function $f_{n,\varrho_n}$, where our FL model converges starting from the 50th epoch. By using Mean Squared Error (MSE) as a loss function, our prediction reaches $0.1119$ MSE.

 Fig. \ref{fig:Fronthaul_Routing} shows comparisons of fronthaul delay using all possible shorted paths and SR paths from ingress TSNBs to egress TSNBs. The results demonstrate that considering the shortest routes and the SR paths, the SR gives the lowest possible latency paths to reach egress TSNBs. 
 After predicting the fronthaul routing and traffic, the  Near-RT RIC shares routing information with MEC server for DQL (joint problem of offloading, fronthaul routing, and edge computation). Furthermore, Fig \ref{fig:fronthaul_delayDQL} shows the advantages of FL by comparing DQL with FL assistance and DQL without FL support. In DQL  assisted by FL, Near-RT RIC shares predicted fronthaul routing information with MEC for DQL. In DQL without FL assistance, near-RT RIC shares fronthaul traffic matrix without using FL for prediction in SR. The simulation results show that DQL performs better when assisted by FL because decisions can be made rapidly for the lowest latency path to reach egress TSNBs. This reduces fronthaul delay.

Fig. \ref{fig:OfflaodingComputation} shows the total delay for computation and offloading when we compute only at edge devices, ECs, or RC.   Computation at edges devices experiences lower latency because local computation does not involve offloading delay. Also, the edge devices can minimize delay by offloading some tasks to ECs or RC.
Furthermore, Fig. \ref{fig:OfflaodingDelay} shows offloading delay where computation delay is excluded in the results. The results from both figures (Figs $11$ and $12$) demonstrate that computation at RC  experiences high delays because the RC is far from edge devices, which involves significant communication latency. Considering the reward function in (\ref{eq:problem_formulation1}), the agent needs to decide to compute locally at edge devices, offload at ECs, or RC to meet computation deadlines and avoid paying penalties. We use $\varpi_\textrm{CoD}=0.5$, $\varpi_\textrm{w}=1\mathrm{e}{-7}$,  $\varpi_\textrm{f}=1\mathrm{e}{-4}$, and $\varpi_\textrm{c}=1\mathrm{e}{-10}$ as penalties. Fig. \ref{fig:Running_Time} shows total computation delay considering all computation and offloading scenarios (at edge devices, ECs, and RC). In other words, in Fig. \ref{fig:Running_Time}, we compared total delay related to local computation,  offloading, routing, and edge computation delays using DQL and Q-learning. This figure clearly shows that using DQL has a minimum delay over Q-learning. In other words, the excellent performance of DQL is thanks to the agent/MEC server that stores previous experiences in local memory and uses neural networks' maximum output to get a new Q-Value. In Fig. \ref{fig:Reward1}, we compare the DQL and Q-learning in terms of reward. We run our simulation for $10000$ episodes. This figure shows that  DQL achieves a better performance than Q-learning in maximizing rewards (i.e., avoiding penalties for missing computation deadlines and violating resource constraints).

We compare DQL in solving (\ref{eq:problem_formulation1}) and optimization-based solution in solving (\ref{eq:problem_formulation0}). To solve (\ref{eq:problem_formulation0}),  we use CVXPY \cite{diamond2016cvxpy} as  Python library for convex optimization problems and BS-MM discussed in Section \ref{sec:Problem_Formulation_Solution}. Using BS-MM for long long-term optimization (considering $10000$ episode corresponds to $10000$ time frames) takes a long time to finish, and always  DQL outperforms BS-MM and Q-learning. In Fig. \ref{fig:Total_delay_BS_MM}, we compute a BS-MM-based solution in a short time frame of $1000$ using Cyclic, Gauss-Southwell, and Randomized indexes selection rules \cite{ndikumana2018joint}.   BS-MM performs better for a short time frame/ short episode, but still, DQL outperforms BS-MM. Therefore, DQL can easily adopt network condition changes for a large episode than BS-MM and Q-learning.

\section{Conclusion}
\label{sec:Conclusion}
Tasks offloading from edge devices to multiple edge clouds requires wireless and fronthaul communication resources to reach edge clouds. Thus,  edge clouds do not have the same available computation resources, and tasks' computation deadlines are different; we need a joint approach for task routing and distribution to multiple edge clouds. This paper proposed a new joint task offloading, segment routing for the fronthaul network, and edge computing approach in O-RAN. We formulated an optimization problem to minimize offloading, routing, and computation delay. We converted the optimization problem to the reward function. Then, we used reinforcement learning and federated learning to maximize the formulated reward by reducing the cost of delay subject to communication and computation resource constraints. The simulation results show that the proposed DQL approach outperforms Q-leaning and BS-MM in minimizing delay and increasing reward. We plan to enhance our offloading and fronthaul routing evaluation using various network scenarios and metrics as future work.
\section*{Acknowledgment}
\noindent The authors thank Mitacs, Ciena, and ENCQOR for funding this research under the IT13947 grant.
\bibliographystyle{IEEEtran}

\begin{thebibliography}{10}
	\bibitem{10045045}
	A.~Ndikumana, K.~K. Nguyen, and M.~Cheriet, ``Federated learning assisted deep
	q-learning for joint task offloading and fronthaul segment routing in open
	ran,'' \emph{IEEE Transactions on Network and Service Management}, pp. 1--1,
	2023.
	
	\bibitem{IoTAnalytics}
	I.~Analytics, ``State of the {IoT} 2018: Number of {IoT} devices now at {7B}
	â€“ market accelerating,''
	\url{https://iot-analytics.com/state-of-the-iot-update-q1-q2-2018-number-of-iot-devices-now-7b/},
	[Online; accessed Jan. 10, 2022].
	
	\bibitem{ndikumana2019joint}
	A.~Ndikumana, N.~H. Tran, T.~M. Ho, Z.~Han, W.~Saad, D.~Niyato, and C.~S. Hong,
	``Joint communication, computation, caching, and control in big data
	multi-access edge computing,'' \emph{IEEE Transactions on Mobile Computing},
	vol.~19, no.~6, pp. 1359--1374, 2019.
	
	\bibitem{allianceORANUseCases}
	O.~R. Alliance, ``{O-RAN} use cases and deployment scenarios,'' \emph{White
		Paper, February}, 2020.
	
	\bibitem{ArchitectureDescription}
	O.~Alliance, ``{O-RAN} architecture description,'' \emph{O-RAN-WG1-O-RAN
		Architecture Description - v01.00.00}, 2020.
	
	\bibitem{ndikumana2022age}
	A.~Ndikumana, K.~K. Nguyen, and M.~Cheriet, ``Age of processing-based data
	offloading for autonomous vehicles in multirats open ran,'' \emph{IEEE
		Transactions on Intelligent Transportation Systems}, 2022.
	
	\bibitem{alimi2017toward}
	I.~A. Alimi, A.~L. Teixeira, and P.~P. Monteiro, ``Toward an efficient c-ran
	optical fronthaul for the future networks: A tutorial on technologies,
	requirements, challenges, and solutions,'' \emph{IEEE Communications Surveys
		\& Tutorials}, vol.~20, no.~1, pp. 708--769, 2017.
	
	\bibitem{farkas2018time}
	J.~Farkas, L.~L. Bello, and C.~Gunther, ``Time-sensitive networking
	standards,'' \emph{IEEE Communications Standards Magazine}, vol.~2, no.~2,
	pp. 20--21, 2018.
	
	\bibitem{2020standard}
	{IEEE}, ``{802.1CMde-2020} - {IEEE} standard for local and metropolitan area
	networks -- time-sensitive networking for fronthaul,''
	\url{https://standards.ieee.org/standard/802_1CM-2018.html}, [Online;
	accessed September. 30, 2020].
	
	\bibitem{perez20195g}
	G.~O. P{\'e}rez, D.~L. Lopez, and J.~A. Hern{\'a}ndez, ``5g new radio fronthaul
	network design for ecpri-ieee 802.1 cm and extreme latency percentiles,''
	\emph{IEEE Access}, vol.~7, pp. 82\,218--82\,230, 2019.
	
	\bibitem{Viavi}
	V.~Solutions, ``{5G} fronthaul handbook,''
	\url{https://gsacom.com/paper/5g-fronthaul-handbook-viavi-solutions-dec-2019/},
	Dec 2019.
	
	\bibitem{tong2021sdn}
	V.~Tong, S.~Souihi, H.~A. Tran, and A.~Mellouk, ``{SDN}-based application-aware
	segment routing for large-scale network,'' \emph{IEEE Systems Journal}, 2021.
	
	\bibitem{carthern2021introduction}
	C.~Carthern, W.~Wilson, and N.~Rivera, ``Introduction to tools and
	automation,'' in \emph{Cisco Networks}.\hskip 1em plus 0.5em minus
	0.4em\relax Springer, 2021, pp. 211--217.
	
	\bibitem{mustafa2021joint}
	E.~Mustafa, J.~Shuja, A.~I. Jehangiri, S.~Din, F.~Rehman, S.~Mustafa,
	T.~Maqsood, A.~N. Khan \emph{et~al.}, ``Joint wireless power transfer and
	task offloading in mobile edge computing: a survey,'' \emph{Cluster
		Computing}, pp. 1--20, 2021.
	
	\bibitem{ndikumana2017collaborative}
	A.~Ndikumana, S.~Ullah, T.~LeAnh, N.~H. Tran, and C.~S. Hong, ``Collaborative
	cache allocation and computation offloading in mobile edge computing,'' in
	\emph{Proceedings of 19th Asia-Pacific Network Operations and Management
		Symposium (APNOMS)}.\hskip 1em plus 0.5em minus 0.4em\relax IEEE, Sep. 27-29,
	2017, pp. 366--369.
	
	\bibitem{ndikumana2019intelligentedge}
	A.~Ndikumana, ``Intelligentedge: Joint communication, computation, caching, and
	control in collaborative multi-access edge computing,'' {K}yung {H}ee
	{U}niversity, 2019.
	
	\bibitem{ndikumana2019self}
	A.~Ndikumana and C.~S. Hong, ``Self-driving car meets multi-access edge
	computing for deep learning-based caching,'' in \emph{Proceedings of 2019
		International Conference on Information Networking (ICOIN)}.\hskip 1em plus
	0.5em minus 0.4em\relax IEEE, Jan. 9-11, 2019 (Kuala Lumpur, Malaysia), pp.
	49--54.
	
	\bibitem{wu2016computing}
	Z.~Wu, K.~Wang, H.~Ji, and V.~C. Leung, ``A computing offloading algorithm for
	f-ran with limited capacity fronthaul,'' in \emph{2016 IEEE International
		Conference on Network Infrastructure and Digital Content (IC-NIDC)}.\hskip
	1em plus 0.5em minus 0.4em\relax IEEE, 2016, pp. 78--83.
	
	\bibitem{kaneva2021multi}
	K.~Kaneva, N.~Aboutorab, and G.~Leu, ``Multi-hop fronthaul offloading in
	learning-aided fog computing,'' in \emph{2021 IEEE 93rd Vehicular Technology
		Conference (VTC2021-Spring)}.\hskip 1em plus 0.5em minus 0.4em\relax IEEE,
	2021, pp. 1--7.
	
	\bibitem{nakayama2018low}
	Y.~Nakayama, D.~Hisano, T.~Kubo, Y.~Fukada, J.~Terada, and A.~Otaka,
	``Low-latency routing scheme for a fronthaul bridged network,'' \emph{Journal
		of Optical Communications and Networking}, vol.~10, no.~1, pp. 14--23, 2018.
	
	\bibitem{park2020communication}
	J.~Park, S.~Samarakoon, A.~Elgabli, J.~Kim, M.~Bennis, S.-L. Kim, and
	M.~Debbah, ``Communication-efficient and distributed learning over wireless
	networks: Principles and applications,'' \emph{arXiv preprint
		arXiv:2008.02608}, 2020.
	
	\bibitem{li2020federated}
	T.~Li, A.~K. Sahu, A.~Talwalkar, and V.~Smith, ``Federated learning:
	Challenges, methods, and future directions,'' \emph{IEEE Signal Processing
		Magazine}, vol.~37, no.~3, pp. 50--60, 2020.
	
	\bibitem{xiong2020intelligent}
	K.~Xiong, S.~Leng, C.~Huang, C.~Yuen, and L.~Guan, ``Intelligent task
	offloading for heterogeneous v2x communications,'' \emph{arXiv preprint
		arXiv:2006.15855}, Jun 29, 2020.
	
	\bibitem{chen2020intelligent}
	J.~Chen, S.~Chen, S.~Luo, Q.~Wang, B.~Cao, and X.~Li, ``An intelligent task
	offloading algorithm (itoa) for uav edge computing network,'' \emph{Digital
		Communications and Networks}, 2020.
	
	\bibitem{chen2020computation}
	X.~Chen, C.~Wu, Z.~Liu, N.~Zhang, and Y.~Ji, ``Computation offloading in beyond
	{5G} networks: A distributed learning framework and applications,''
	\emph{arXiv preprint arXiv:2007.08001}, 15 Jul. 2020.
	
	\bibitem{rantakokko2022data}
	S.~Rantakokko, ``Data handling process in extended reality ({XR}) when
	delivering technical instructions,'' \emph{Technical Communication}, vol.~69,
	no.~2, pp. 75--96, 2022.
	
	\bibitem{nguyen2021mobile}
	T.~N. Nguyen and S.~Zeadally, ``Mobile crowd-sensing applications: Data
	redundancies, challenges, and solutions,'' \emph{ACM Transactions on Internet
		Technology (TOIT)}, vol.~22, no.~2, pp. 1--15, 2021.
	
	\bibitem{huang2020prospect}
	Y.~Huang, X.~Xu, N.~Li, H.~Ding, and X.~Tang, ``Prospect of 5g intelligent
	networks,'' \emph{IEEE Wireless Communications}, vol.~27, no.~4, pp. 4--5,
	2020.
	
	\bibitem{zhou2017resource}
	Y.~Zhou, F.~R. Yu, J.~Chen, and Y.~Kuo, ``Resource allocation for
	information-centric virtualized heterogeneous networks with in-network
	caching and mobile edge computing,'' \emph{IEEE Transactions on Vehicular
		Technology}, vol.~66, no.~12, pp. 11\,339--11\,351, 2017.
	
	\bibitem{hao2016optimizing}
	F.~Hao, M.~Kodialam, and T.~Lakshman, ``Optimizing restoration with segment
	routing,'' in \emph{IEEE INFOCOM 2016-The 35th Annual IEEE International
		Conference on Computer Communications}.\hskip 1em plus 0.5em minus
	0.4em\relax IEEE, 2016, pp. 1--9.
	
	\bibitem{trimponias2017centrality}
	G.~Trimponias, Y.~Xiao, H.~Xu, X.~Wu, and Y.~Geng, ``Centrality-based
	middlepoint selection for traffic engineering with segment routing,''
	\emph{arXiv preprint arXiv:1703.05907}, 2017.
	
	\bibitem{tun2019weighted}
	Y.~K. Tun, S.~R. Pandey, M.~Alsenwi, C.~W. Zaw, and C.~S. Hong, ``Weighted
	proportional allocation based power allocation in wireless network
	virtualization for future wireless networks,'' in \emph{2019 International
		Conference on Information Networking (ICOIN)}.\hskip 1em plus 0.5em minus
	0.4em\relax IEEE, 2019, pp. 284--289.
	
	\bibitem{sun2016majorization}
	Y.~Sun, P.~Babu, and D.~P. Palomar, ``Majorization-minimization algorithms in
	signal processing, communications, and machine learning,'' \emph{IEEE
		Transactions on Signal Processing}, vol.~65, no.~3, pp. 794--816, 2016.
	
	\bibitem{ndikumana2020deep}
	A.~Ndikumana, N.~H. Tran, K.~T. Kim, C.~S. Hong \emph{et~al.}, ``Deep learning
	based caching for self-driving cars in multi-access edge computing,''
	\emph{IEEE Transactions on Intelligent Transportation Systems}, Mar 4, 2020.
	
	\bibitem{sutton2018reinforcement}
	R.~S. Sutton and A.~G. Barto, \emph{Reinforcement learning: An
		introduction}.\hskip 1em plus 0.5em minus 0.4em\relax MIT press, 2018.
	
	\bibitem{feinberg2012handbook}
	E.~A. Feinberg and A.~Shwartz, \emph{Handbook of Markov decision processes:
		methods and applications}.\hskip 1em plus 0.5em minus 0.4em\relax Springer
	Science \& Business Media, 2012, vol.~40.
	
	\bibitem{fan2020theoretical}
	J.~Fan, Z.~Wang, Y.~Xie, and Z.~Yang, ``A theoretical analysis of deep
	q-learning,'' in \emph{Learning for Dynamics and Control}.\hskip 1em plus
	0.5em minus 0.4em\relax PMLR, 2020, pp. 486--489.
	
	\bibitem{haddadpour2019convergence}
	F.~Haddadpour and M.~Mahdavi, ``On the convergence of local descent methods in
	federated learning,'' \emph{arXiv preprint arXiv:1910.14425}, 2019.
	
	\bibitem{NetworkX}
	{NetworkX}, ``{NetworkX}: Network analysis with python,''
	\url{https://networkx.org/}, [Online; accessed Nov. 22, 2021].
	
	\bibitem{paszke2019pytorch}
	A.~Paszke, S.~Gross, F.~Massa, A.~Lerer, J.~Bradbury, G.~Chanan, T.~Killeen,
	Z.~Lin, N.~Gimelshein, L.~Antiga \emph{et~al.}, ``Pytorch: An imperative
	style, high-performance deep learning library,'' \emph{Advances in neural
		information processing systems}, vol.~32, pp. 8026--8037, 2019.
	
	\bibitem{brockman2016openai}
	G.~Brockman, V.~Cheung, L.~Pettersson, J.~Schneider, J.~Schulman, J.~Tang, and
	W.~Zaremba, ``Openai gym,'' \emph{arXiv preprint arXiv:1606.01540}, 2016.
	
	\bibitem{TensorFlow}
	{T}ensor{F}low, ``{TensorFlow} {Federated}: Machine learning on decentralized
	data,'' \url{https://www.tensorflow.org/federated}, [Online; accessed
	September. 30, 2020].
	
	\bibitem{alom2018history}
	M.~Z. Alom, T.~M. Taha, C.~Yakopcic, S.~Westberg, P.~Sidike, M.~S. Nasrin,
	B.~C. Van~Esesn, A.~A.~S. Awwal, and V.~K. Asari, ``The history began from
	alexnet: A comprehensive survey on deep learning approaches,'' \emph{arXiv
		preprint arXiv:1803.01164}, 2018.
	
	\bibitem{wei2021shortest}
	H.~Wei, S.~Zhang, and X.~He, ``Shortest path algorithm in dynamic restricted
	area based on unidirectional road network model,'' \emph{Sensors}, vol.~21,
	no.~1, p. 203, 2021.
	
	\bibitem{diamond2016cvxpy}
	S.~Diamond and S.~Boyd, ``{CVXPY}: {A} {P}ython-embedded modeling language for
	convex optimization,'' \emph{Journal of Machine Learning Research}, vol.~17,
	no.~83, pp. 1--5, 2016.
	
	\bibitem{ndikumana2018joint}
	A.~Ndikumana, N.~H. Tran, T.~M. Ho, D.~Niyato, Z.~Han, and C.~S. Hong, ``Joint
	incentive mechanism for paid content caching and price based cache
	replacement policy in named data networking,'' \emph{IEEE Access}, vol.~6,
	pp. 33\,702--33\,717, 2018.
	
\end{thebibliography}

\begin{IEEEbiography}[{\includegraphics[width=1in,height=1.25in,clip,keepaspectratio]{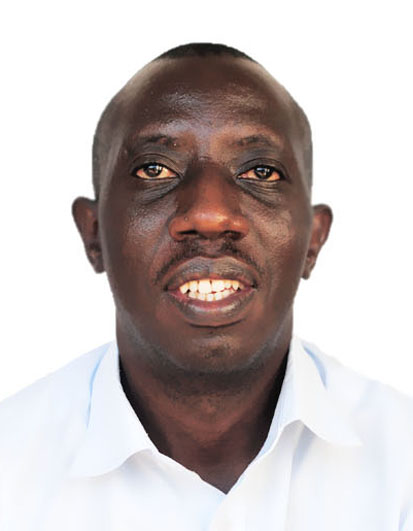}}]{Anselme Ndikumana} received  B.S. degree in Computer Science from the National University of Rwanda in 2007 and Ph.D. degree in Computer Engineering from Kyung Hee University, South Korea in August 2019. Since 2020, he has been	with the Synchromedia Lab, École de Technologie Supérieure, Université du Québec, Montréal, QC, Canada, where he is currently a Postdoctoral Researcher. His professional experience includes Lecturer at the University of Lay Adventists of Kigali from 2019 to 2020, Chief Information System, a System Analyst, and a Database Administrator at Rwanda Utilities Regulatory Authority from 2008 to 2014. His research interest includes AI for wireless communication, multi-access edge computing, 5G and next-G networks, metaverse, network economics, game theory,  and network optimization.
\end{IEEEbiography}
\begin{IEEEbiography}[{\includegraphics[width=1in,height=1.25in,clip,keepaspectratio]{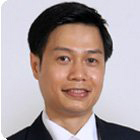}}]{Kim Khoa Nguyen} is Associate Professor in the Department of Electrical Engineering and the founder of the Laboratory on IoT and Cloud Computing at the University of Quebec’s  Ecole de technologie superieure. In the past, he served as CTO of Inocybe Technologies (now is Kontron Canada), a world’s leading company in software-defined networking (SDN) solutions. He was the architect of the Canarie’s GreenStar Network and led R\&D in large-scale projects with Ericsson, Ciena, Telus, InterDigital, and Ultra Electronics. He is the recipient of Microsoft Azure Global IoT Contest Award 2017, and Ciena’s Aspirational Prize 2018. He is the author of more than 100 publications, and holds several industrial patents. His expertise includes network optimization, cloud computing IoT, 5G, big data, machine learning, smart city, and high speed networks.
\end{IEEEbiography}
\begin{IEEEbiography}[{\includegraphics[width=1in,height=1.25in,clip,keepaspectratio]{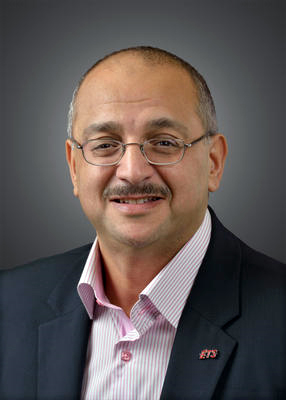}}]{Dr. Mohamed Cheriet} received his Bachelor, M.Sc. and Ph.D. degrees in Computer Science from USTHB (Algiers) and the University of Pierre \& Marie Curie (Paris VI) in 1984, 1985 and 1988 respectively. He was then a Postdoctoral Fellow at CNRS, Pont et Chaussées, Paris V, in 1988, and at CENPARMI, Concordia U., Montreal, in 1990. Since 1992, he has been a professor in the Systems Engineering department at the University of Quebec - École de Technologie Supérieure (ÉTS), Montreal, and was appointed full Professor there in 1998. Prof. Cheriet was the director of LIVIA Laboratory for Imagery, Vision, and Artificial Intelligence (2000-2006), and is the founder and director of Synchromedia Laboratory for multimedia communication in telepresence applications, since 1998.  Dr. Cheriet research has extensive experience in Sustainable and Intelligent Next Generation Systems. Dr. Cheriet is an expert in Computational Intelligence, Pattern Recognition, Machine Learning, Artificial Intelligence and Perception and their applications, more extensively in Networking and Image Processing. In addition, Dr. Cheriet has published more than 500 technical papers in the field and serves on the editorial boards of several renowned journals and international conferences. He held a Tier 1 Canada Research Chair on Sustainable and Smart Eco-Cloud (2013-2000), and lead the establishment of the first smart university campus in Canada, created as a hub for innovation and productivity at Montreal. Dr. Cheriet is the General Director of the FRQNT Strategic Cluster on the Operationalization of Sustainability Development, CIRODD (2019-2026). He is the Administrative Director of the \$12M CFI’2022 CEOS*Net Manufacturing Cloud Network. He is a 2016 Fellow of the International Association of Pattern Recognition (IAPR), a 2017 Fellow of the Canadian Academy of Engineering (CAE), a 2018 Fellow of the Engineering Institute of Canada (EIC), and a 2019 Fellow of Engineers Canada (EC). Dr. Cheriet is the recipient of the 2016 IEEE J.M. Ham Outstanding Engineering Educator Award, the 2013 ÉTS Research Excellence prize, for his outstanding contribution in green ICT, cloud computing, and big data analytics research areas, and the 2012 Queen Elizabeth II Diamond Jubilee Medal. He is a senior member of the IEEE, the founder and former Chair of the IEEE Montreal Chapter of Computational Intelligent Systems (CIS), a Steering Committee Member of the IEEE Sustainable ICT Initiative, and the Chair of ICT Emissions Working Group. He contributed 6 patents (3 granted), and the first standard ever, IEEE 1922.2, on real-time calculation of ICT emissions, in April 2020, with his IEEE Emissions Working Group.
\end{IEEEbiography}
\end{document}